\documentclass{report}
\usepackage{amssymb}

\usepackage{amsmath}
\usepackage{graphicx}

\input{tcilatex}

\begin{document}

\title{{\huge Quantum Information Theory}\\
{\huge and}\\
{\huge Applications to Quantum Cryptography} \vspace{3.2cm}}
\author{By \vspace{0.4cm} \\
%EndAName
\textbf{Nikolaos P. Papadakos}\\
\emph{B.Sc. in Mathematics, University of Athens, 2001} \vspace{2cm}\\
Supervisor\\
\textbf{Professor Abbas Edalat \vspace{3.2cm}}\\
{\large Individual Study Option,}\\
{\large on the First Term of the Academic year 2001-2002,}\\
{\large for the M.Sc. in Advanced Computing\vspace{3.2cm}}\\
{\Large \emph{Department of Computing}}\\
{\Large \emph{Imperial College of Science, Technology and Medicine}}\\
{\Large \emph{University of London, United Kingdom} \pagenumbering{roman}}\\
}
\date{}
\maketitle
\tableofcontents
\listoffigures

\addtocontents{toc}{\contentsline{chapter}{List of Figures}{iii}}

\chapter*{Acknowledgements\label{chap_Acknowledgements}}

\addtocontents{toc}{\contentsline{chapter}{Acknowledgements}
{\pageref{chap_Acknowledgements}}}

I am very pleased to thank my supervisor Professor Abbas Edalat, who with
his important comments not only did he help me improve the presentation of
this Individual Study Option, but by suggesting me to solve all 64 exercises
of the last two chapters of \cite[chap.11,12]{NielsenChuang} he contributed
dramatically to my understanding and augmented my self-confidence on the
field. I am indebted to him, for his effort of transforming the present
material into something more than just an Individual Study Option.

\chapter*{About this Individual Study Option\label{chap_About}}

\addtocontents{toc}
{\contentsline{chapter}{About this Individual Study Option}
{\pageref{chap_About}}}

In this Individual Study Option the concepts of classical and quantum
information theory are presented, with some real-life applications, such as
data compression, information transmission over noisy channels and quantum
cryptography.

\subsubsection{Abstract}

In each chapter the following topics are covered:

\begin{enumerate}
\item The relation of information theory with physics is studied and by
using the notion of Shannon's entropy, measures of information are
introduced. Then their basic properties are proven. Moreover important
information relevant features of quantum physics, like the disability of
distinguishing or cloning states in general are studied. In order to get a
quantum measure of information, an introduction to quantum thermodynamics is
given with a special focus on the explanation of the utility of density
matrix. After this von Neumann entropy and its related measures are defined.
In this context a major discrepancy between classical and quantum
information theory is presented: quantum entanglement. The basic properties
of von Neumann entropy are proven and some information theoretic
interpretation of quantum measurement is given.

\item The amount of accessible information is obtained by the Fano's
inequality in the classical case, and by its quantum analogue and Holevo's
bound in the quantum case. In addition to this classical and quantum data
processing is discussed.

\item Furthermore for real-life applications data compression is studied via
Shannon's classical and Schumacher's quantum noiseless channel coding
theorems. Another application is transmission of classical information over
noisy channels. For this a summary of classical and quantum error correction
is given, and then Shannon's classical noisy channel and
Holevo-Schumacher-Westmoreland quantum noisy channel coding theorems are
studied. The present state of transmission of quantum information over
quantum channels is summarized.

\item A practical application of the aforementioned is presented: quantum
cryptography. In this context the BB84, a quantum key distribution protocol,
is demonstrated and its security is discussed. The current experimental
status of quantum key distribution is summarized and the possibility of a
commercial device realizing quantum cryptography is presented.
\end{enumerate}

\subsubsection{Special viewpoints emphasized}

\begin{figure}[tbh]
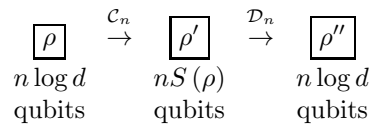

\begin{center}
\begin{equation*}
\underset{%
\begin{array}{c}
n\log d \\ 
\text{qubits}%
\end{array}%
}{\fbox{$\rho $}}\overset{\mathcal{C}_{n}}{\overset{}{\rightarrow }}\underset%
{%
\begin{array}{c}
nS\left( \rho \right) \\ 
\text{qubits}%
\end{array}%
}{\fbox{$\rho ^{\prime }$}}\overset{\mathcal{D}_{n}}{\overset{}{\rightarrow }%
}\underset{%
\begin{array}{c}
n\log d \\ 
\text{qubits}%
\end{array}%
}{\fbox{$\rho ^{\prime \prime }$}}
\end{equation*}%
\end{center}
\caption{Quantum data compression. The compression operation $\mathcal{C}%
_{n} $ compresses a quantum source $\protect\rho $ stored in $n\log d$
qubits into $nS\left( \protect\rho \right) $ qubits. The source is
accurately recovered via the decompression operation $\mathcal{D}_{n}.$}
\label{fig_Schumacher_coding}
\end{figure}

The above material is in close connection with the last two chapters of %
\cite[chap.11,12]{NielsenChuang}, and error correction is a summary of the
results given in chapter 10 of \cite{NielsenChuang}. From this book Figures %
\ref{fig_Entropy_diagram}, \ref{fig_Schumacher_coding}, \ref%
{fig_Noisy_channel_coding} and \ref{fig_Sum_Info_Theory}, were extracted.
However there was much influence on information theory by \cite{Preskill},
and some results given therein like for example equation (\ref%
{Shannon_data_compress_intuitive_proof_for_2}) were extended to (\ref%
{Shannon_data_compress_intuitive_proof_for_n}). Of equal impact was \cite%
{Steane97} concerning error correction, and \cite{Steane97,Gisin-et-al}
concerning quantum cryptography. Note that Figure was \ref{fig_QKD}\
extracted from \cite{Gisin-et-al}. Of course some notions which were vital
to this Individual Study Option were taken from almost all the chapters of %
\cite{NielsenChuang}.

However there where some topics not well explained or sometimes not enough
emphasized. The most important tool of quantum information theory, the
density matrix, is misleadingly given in \cite[p.98]{NielsenChuang}: \emph{%
''the density matrix [...] is mathematically equivalent to the state vector
approach, but it provides a much more convenient language for thinking about
commonly encountered scenarios in quantum mechanics''}. The density matrix
is not equivalent to the state vector approach and is much more than just a
convenient language. The former is a tool for quantum thermodynamics, a
field where is impossible to use the latter. Rephrasing it, quantum
thermodynamics is the only possible language. Willing to augment
understanding of this meaningful tool, subsection \ref%
{subsec_Quantum_thermodynamics} is devoted to it, and is close to the
viewpoint of \cite[p.295-307]{Cohen-Tanoudji}, which is a classical textbook
for quantum mechanics.

In this Individual Study Option some topics are presented in a different
perspective. There is an attempt to emphasize information theoretic
discrepancies between classical and quantum information theory, the greatest
of which is perhaps quantum entanglement. A very special demonstration of
this difference is given in subsection \ref{subsec_QuantumEntanglement}. It
should be noted that some particular information theoretic meaning of
measurement in physics is presented in subsection \ref%
{subsec_Measurement_and_Info}.

Concerning the different views presented in this text, nothing could be more
beneficial than the 64 exercises of the last two chapters of %
\cite[chap.11,12]{NielsenChuang}. As an example a delicate presentation of
measures of information is given in page \pageref%
{presentation_of_information_measures}, due to exercise 11.2 \cite[p.501]%
{NielsenChuang}, or subsection \ref{subsec_QuantumEntanglement} on quantum
entanglement was inspired by exercise 11.14 \cite[p.514]{NielsenChuang}. In
some occasions special properties where proved in order to solve the
exercises. Such a property is the preservation of quantum entropy by unitary
operations (property 2 of von Neumann entropy, in page \pageref%
{Unitary_operation_preserve_entropy}), which was needed to solve for example
exercise 11.19,20 \cite[p.517,518]{NielsenChuang}. For these last two
exercises some special lemmas were proven in appendices \ref%
{appendix_Special_diagonal_normal_matr} and \ref%
{appendix_Proj_measur_and_Unitary}. It is important to notice that from the
above mentioned exercises of \cite[chap.11,12]{NielsenChuang} only half of
them are presented here.

\subsubsection{Notation}

Finally about the mathematical notation involved, the symbol $\triangleq $
should be interpreted as \emph{''defined by''}, and the symbol $\equiv $ as 
\emph{''to be identified with''}.

\chapter{Physics and information theory}

\setcounter{page}{1} \pagenumbering{arabic}

Since its invention, computer science \cite{Turing}\ was considered as a
branch of mathematics, in contrast to information theory \cite%
{Shannon,ShannonWeaver} which was viewed by its physical realization;
quoting Rolf Landauer \cite{Landauer} \emph{''information is physical''}.
The last decades changed the landscape and both computers and information
are mostly approached by their physical implementations \cite%
{Manin,Benioff,Feynman,Deutsch}. This view is not only more natural, but in
the case of quantum laws it gives very exciting results and sometimes an
intriguing view of what information is or can be. Such an understanding
could never be inspired just from mathematics. Moreover there is a
possibility that the inverse relation exists between physics and information %
\cite{IngardenUrbanik(61),IngardenUrbanik(62)}\ or quoting Steane \cite%
{Steane97} one could find a new methods of studying physics by \emph{''the
ways that nature allows, and prevents, information to be expressed and
manipulated, rather than [}the ways it allows\emph{] particles to move''}.
Such a program is still in its infancy, however one relevant application is
presented in subsection \ref{subsec_Measurement_and_Info}. In this chapter
the well established information theory based on classical physics is
presented in section \ref{sec_Classical_physics_&_info_th}, and the
corresponding up to date known results for quantum physics are going to be
analyzed in section \ref{sec_Quant_phys_and_info_theory}.

\section{Classical physics and information theory\label%
{sec_Classical_physics_&_info_th}}

In classical physics all entities have certain properties which can be known
up to a desired accuracy. This fact gives a simple pattern to store and
transmit information, by assigning information content to each of the
property a physical object can have. For example storage can be realized by
writing on a paper, where information lays upon each letter, or on a
magnetic disk, where information, in binary digits (bits), is represented
each of the two spin states a magnetic dipole can have. In what concerns
transmission, speech is one example, where each sound corresponds to an
information, or a second example is an electronic signal on a wire, where
each state of the electricity is related to some piece of information.
Unfortunately in every day life such simple patterns are non-economical and
unreliable. This is because communication is realized by physical entities
which are imperfect, and hence they can be influenced by environmental
noise, resulting information distortion.

Concerning the economical transmission of information, one can see that the
naive pattern of information assignment presented in the last paragraph is
not always an optimal choice. This is because a message, in English language
for example, contains symbols with different occurrence frequencies. Looking
for example this text one can note immediately that the occurrence
probability of letter \texttt{a}, $p_{\mathtt{a}},$ is much greater than
that of exclamation $p_{\text{\texttt{!}}}.$ According to the naive
assignment, English language symbols are encoded to codewords of identical
length $l$, and the average space needed to store $\mathtt{a}$\ is $lp_{%
\mathtt{a}}$ and of the exclamation $lp_{\text{\texttt{!}}},$ and since $p_{%
\mathtt{a}}>p_{\text{\texttt{!}}}$, a lot of space is wasted for the letter $%
\mathtt{a}$. In order to present how a encoding scheme can be economical
consider a four letter alphabet \texttt{A}, \texttt{B}, \texttt{C}, \texttt{D%
}, with occurrence probabilities $p_{\mathtt{A}}=\frac{3}{4},$ $p_{\mathtt{B}%
}=\frac{1}{8},$ $p_{\mathtt{C}}=p_{\mathtt{D}}=\frac{1}{16},$ the subsequent
assignment of bits: $\mathtt{A}\rightarrow 1,$ $\mathtt{B}\rightarrow 01,$ $%
\mathtt{C}\rightarrow 010,$ and $\mathtt{D}\rightarrow 011$. A message of $n$
symbols, using this encoding, has on average $n\left( p_{\mathtt{A}}+2p_{%
\mathtt{B}}+3p_{\mathtt{C}}+3p_{\mathtt{D}}\right) =n\left( \frac{3}{4}+2%
\frac{1}{8}+3\frac{1}{16}+3\frac{1}{16}\right) =n\left( \frac{11}{8}\right) $
bits instead of $2n$ which would needed if somebody just mapped to each
letter a two bit codeword.

The topics discussed in the last two paragraphs give rise to the most
important information theoretic question: \emph{which are the minimal
resources needed to reliably communicate?} An answer to this question can be
given by abstractly quantifying information in relevance to the physical
resources needed to carry it. Motivated by the previously demonstrated four
letter alphabet example, probabilities are going to be used for such an
abstraction. One now defines a function $H,$ quantifying a piece of
information $I,$ exhibiting the following reasonable properties:\label%
{presentation_of_information_measures}

\begin{enumerate}
\item $H\left( I\right) $ is a function only of the probability of
occurrence $p$ of information $I,$ thus $H\left( I\right) \equiv H\left(
p\right) .$

\item $H$ is a smooth function.

\item The resources needed for two independent informations with individual
probabilities $p,q>0\ $are the sum of the resources needed for one alone, or
in mathematical language $H\left( pq\right) =H\left( p\right) +H\left(
q\right) .$
\end{enumerate}

The second and third property imply that $H$ is a logarithmic function, and
by setting $q=1$ in the third it is immediate to see that $H\left( 1\right)
=0.$ Hence $H\left( p\right) =k\log p,$ where $k$ and $a$ are constants to
be determined (refer to comments after equations (\ref{def_entropy}) and (%
\ref{entopy<_relation})). This means that the average of resources needed
when one of the mutually exclusive set of information with probabilities $%
p_{1},\ldots ,p_{n}$ occurs is 
\begin{equation}
H\left( p_{1},\ldots ,p_{n}\right) =k\sum_{i}p_{i}\log _{a}p_{i}.
\label{intuitive_entropy}
\end{equation}%
It should be noted that probability is not the only way of quantifying
information \cite%
{IngardenUrbanik(61),IngardenUrbanik(62),Ingarden(63),Ingarden(64),Ingarden(65),Urbanik(72)}%
.

\subsection{Shannon entropy and relevant measures of classical information
theory}

The function $Q$ found in (\ref{intuitive_entropy}) is known in physics as 
\emph{entropy}, and measures the order of a specific statistical system. Of
course one interesting physical system is an $n$-bit computer memory, and if
all the possible cases of data entry are described by a random variable $X$
with probabilities $p_{1},\ldots ,p_{2^{n}},$ then the computer's memory
should have an entropy given by%
\begin{equation}
H\left( X\right) \equiv H\left( p_{1},\ldots ,p_{2^{n}}\right) \triangleq -%
\underset{x}{\sum }p_{x}\log p_{x}.  \label{def_entropy}
\end{equation}%
Here a modified version of (\ref{intuitive_entropy}) is used, with $\log
\equiv \log _{2}$ and $k\triangleq 1,$ an assignment to be verified after
equation (\ref{entopy<_relation}). Equation (\ref{def_entropy}) is known in
information theory as the \emph{Shannon's entropy} \cite{Shannon}. There are
two complementary ways of understanding Shannon's entropy. It can be
considered as a measure of uncertainty before learning the value of a
physical information or the information gain after learning it.

The Shannon entropy gives rise to other important measures of information.
One such is the \emph{relative entropy}, which is defined by%
\begin{equation}
H\left( p_{x}||q_{x}\right) \triangleq \underset{x}{\sum }p_{x}\log \frac{%
p_{x}}{q_{x}}=-H\left( X\right) -\underset{x}{\sum }p_{x}\log q_{x},
\label{def_relative_entropy}
\end{equation}%
and is a measure of distance between two probability distributions. This is
because it can be proven that%
\begin{eqnarray}
H\left( p_{x}||q_{x}\right) &>&0,  \label{rel_entropy_metric_like_properties}
\\
H\left( p_{x}||q_{x}\right) &=&0\Leftrightarrow \forall x:p_{x}=q_{x}. 
\notag
\end{eqnarray}%
Of course it is not a metric because as one can check $H\left(
p_{x}||q_{x}\right) =H\left( q_{x}||p_{x}\right) $ is not always true. The
relative entropy is often useful, not in itself, but because it helps
finding important results. One such is derived using the last equality in (%
\ref{def_relative_entropy}) and (\ref{rel_entropy_metric_like_properties}),
then in a memory with $n$-bits%
\begin{eqnarray}
H\left( X\right) &<&\log 2^{n}=n,  \label{entopy<_relation} \\
H(X) &=&\log 2^{n}=n\Leftrightarrow \forall i:p_{i}=\frac{1}{2^{n}},  \notag
\end{eqnarray}%
which justifies the selection of $k=1$ in (\ref{def_entropy}), because in
the optimal case and in absence of noise, the maximum physical resources
needed to transmit or to store an $n$-bit word should not exceed $n$. One
can also see from (\ref{def_entropy}) that%
\begin{eqnarray}
H\left( X\right) &>&0,  \label{entropy>_relation} \\
H\left( X\right) &=&0\Leftrightarrow \text{system }X\text{ is in a definite
state (}p=1\text{).}  \notag
\end{eqnarray}

Other important results are deduced relative entropy, and concern useful
entropic quantities such as the \emph{joint entropy}, the \emph{entropy of }$%
X$\emph{\ conditional on knowing }$Y,$ and the \emph{common} or \emph{mutual
information of }$X$\emph{\ and }$Y.$ Those entropies are correspondingly
defined by the subsequent intuitive relations%
\begin{eqnarray}
H\left( X,Y\right) &\triangleq &-\sum_{xy}p_{xy}\log p_{xy},
\label{def_Shan_entr_joint} \\
H\left( X|Y\right) &\triangleq &H\left( X,Y\right) -H\left( Y\right) ,
\label{def_Shan_entr_conditional} \\
H\left( X:Y\right) &\triangleq &H\left( X\right) +H\left( Y\right) -H\left(
X,Y\right) =H\left( X\right) -H\left( X|Y\right) ,
\label{def_Shan_entr_mutual}
\end{eqnarray}%
and can be represented in the 'entropy Venn diagram' as shown in Figure \ref%
{fig_Entropy_diagram}.

\begin{figure}[tbh]
\centering \includegraphics[width=8cm]{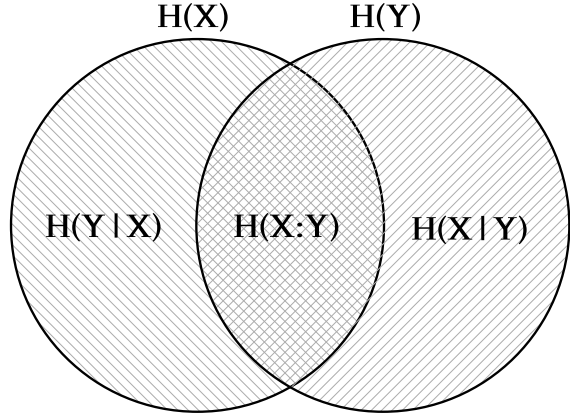}
\caption{Relationships between different entropies.}
\label{fig_Entropy_diagram}
\end{figure}

\subsection{Basic properties of Shannon entropy\label%
{sec_BasicPropertiesofShannonEntropy}}

It is worth mentioning here the basic properties of Shannon entropy:

\begin{enumerate}
\item $H\left( X,Y\right) =H\left( Y,X\right) ,$ $H\left( X:Y\right)
=H\left( Y:X\right) .$

\item $H\left( Y|X\right) \geq 0$ and thus by the second equality of (\ref%
{def_Shan_entr_mutual}) $H\left( X:Y\right) \leq H\left( Y\right) ,$ with
equality if and only if $Y$ is a function of $X,$ $Y=f\left( X\right) .$

\item $H\left( X\right) \leq H\left( X,Y\right) ,$ with equality if and only
if $Y$ is a function of $X.$

\item \textbf{Subadditivity:} $H\left( X,Y\right) \leq H\left( X\right)
+H\left( Y\right) $ with equality if and only if $\ X$ and $\ Y$ are
independent random variables.

\item $H\left( Y|X\right) \leq H\left( Y\right) $ and thus by the second
equality of (\ref{def_Shan_entr_mutual}) $H\left( X:Y\right) \geq 0,$ with
equality in each if and only if $X$ and $Y$ are independent variables.

\item \textbf{Strong subadditivity:} $H\left( X,Y,Z\right) +H\left( X\right)
\leq H\left( X,Y\right) +H\left( Y,Z\right) ,$ with equality if and only if $%
Z\rightarrow Y\rightarrow X$ forms a Markov chain.

\item \textbf{Conditioning reduces entropy:} $H\left( X|Y,Z\right) \leq
H\left( X|Y\right) .$

\item \textbf{Chaining rule for conditional entropies:} Let $X_{1},\ldots
,X_{n}$ and $Y$ be any set of random variables, then $H\left( X_{1},\ldots
,X_{n}|Y\right) =\sum_{i=1}^{n}H\left( X_{i}|Y,X_{1},\ldots ,X_{i-1}\right)
. $

\item \textbf{Concavity of the entropy:} Suppose there are probabilities $%
q_{i}\geq 0,$ $p_{i}>0$ and then $H\left( \sum_{i}p_{i}q_{j}\right) \geq
\sum_{i}p_{i}H\left( q_{j}\right) ,$ with equality if and only if $q_{j}$s
are identical.
\end{enumerate}

The various relationships between entropies may mostly be derived from the
'entropy Venn diagram' shown in Figure \ref{fig_Entropy_diagram}. Such
figures are not completely reliable as a guide to properties of entropy, but
they provide a useful mnemonic for remembering the various definitions and
properties of entropy. The proofs of the above mentioned properties follow.

\emph{Proof}

\begin{enumerate}
\item Obvious from the definition of joint entropy (\ref{def_Shan_entr_joint}%
) and mutual entropy (\ref{def_Shan_entr_mutual}).

\item Based on the definition of conditional entropy (\ref%
{def_Shan_entr_conditional}) 
\begin{eqnarray*}
H\left( Y|X\right) &\triangleq &H\left( X,Y\right) -H\left( Y\right)
=-\sum_{xy}p\left( x,y\right) \log p\left( x,y\right) +\sum_{x}p\left(
y\right) \log p\left( y\right) \\
&=&-\sum_{xy}p\left( x,y\right) \log p\left( x,y\right) +\sum_{xy}p\left(
x,y\right) \log p\left( y\right) =-\sum_{xy}p\left( x,y\right) \log \frac{%
p\left( x,y\right) }{p\left( y\right) } \\
&=&-H\left( p\left( x,y\right) ||p\left( y\right) \right) ,
\end{eqnarray*}%
and using (\ref{rel_entropy_metric_like_properties}), where equality holds
if and only if $p\left( X,Y\right) =p\left( Y\right) ,$ that is $Y=f\left(
X\right) .$

\item It is immediately proven by the second one and using the definition of
conditional entropy (\ref{def_Shan_entr_conditional}).

\item Following the subsequent steps of calculation%
\begin{eqnarray*}
H\left( p\left( x,y\right) ||p\left( x\right) p\left( y\right) \right)
&\triangleq &-H\left( p\left( x,y\right) \right) -\sum_{xy}p\left(
x,y\right) \log p\left( x\right) p\left( y\right) \\
&=&-H\left( p\left( x,y\right) \right) -\sum_{xy}p\left( x,y\right) \log
p\left( x\right) -\sum_{xy}p\left( x,y\right) \log p\left( y\right) \\
&=&-H\left( p\left( x,y\right) \right) +H\left( p\left( x\right) \right)
+H\left( p\left( y\right) \right) ,
\end{eqnarray*}%
where the second equality of (\ref{def_relative_entropy}) was used. The
result follows directly from equation (\ref%
{rel_entropy_metric_like_properties}).

\item It is easily derived from subadditivity (property 4) and the
definition of conditional entropy (\ref{def_Shan_entr_conditional}).

\item First note that by the definition of the joint entropy (\ref%
{def_Shan_entr_joint}) and some algebra $H\left( X,Y,Z\right) +H\left(
Y\right) -H\left( X,Y\right) -H\left( Y,Z\right) =\sum_{x,y,z}p\left(
x,y,z\right) \log \frac{p\left( x,y\right) p\left( y,z\right) }{p\left(
x,y,z\right) p\left( y\right) }.$ Then using the fact that $\log x\leq \frac{%
x-1}{\ln 2}$ for all positive $x$ and equality achieved if and only if $x=1,$
the following can be concluded%
\begin{eqnarray*}
\sum_{xyz}p\left( x,y,z\right) \log \frac{p\left( x,y\right) p\left(
y,z\right) }{p\left( x,y,z\right) p\left( y\right) } &\leq &\sum_{xyz}\frac{%
p\left( x,y,z\right) }{\ln 2}\left( \frac{p\left( x,y\right) p\left(
y,z\right) }{p\left( x,y,z\right) p\left( y\right) }-1\right) \\
&=&\frac{1}{\ln 2}\left( \sum_{y}\frac{p\left( y\right) p\left( y\right) }{%
p\left( y\right) }\right) -\frac{1}{\ln 2}=0,
\end{eqnarray*}%
with equality if and only if $\frac{p\left( x,y\right) p\left( y,z\right) }{%
p\left( x,y,z\right) p\left( y\right) }=1\Leftrightarrow p\left( z|y\right)
=p\left( z|x,y\right) \Leftrightarrow X\rightarrow Y\rightarrow Z$ is a
Markov chain, q.e.d.

\item From strong subadditivity (property 6) it straightforward that $%
H\left( X,Y,Z\right) -H\left( Y,Z\right) \leq H\left( X,Y\right) -H\left(
Y\right) $ and from the definition of conditional entropy (\ref%
{def_VonNeum_conditional}) the result is obvious.

\item First the result is proven for $n=2$ using the definition of
conditional entropy (\ref{def_VonNeum_conditional})%
\begin{eqnarray*}
H\left( X_{1},X_{2}|Y\right) &=&H\left( X_{1},X_{2},Y\right) -H\left(
Y\right) =H\left( X_{1},X_{2},Y\right) -H\left( X_{1},Y\right) +H\left(
X_{1},Y\right) -H\left( Y\right) \\
&=&H\left( X_{2}|Y,X_{1}\right) +H\left( X_{1}|Y\right) .
\end{eqnarray*}%
Now induction is going to be used to prove it for every $n$. Assume that the
result holds for $n,$ then using the one for $n=2,$ $H\left( X_{1},\ldots
,X_{n+1}|Y\right) =H\left( X_{2},\ldots ,X_{n+1}|Y,X_{1}\right) +H\left(
X_{1}|Y\right) ,$ and applying the inductive hypothesis to the first term on
the right hand side gives%
\begin{equation*}
H\left( X_{1},\ldots ,X_{n+1}|Y\right) =\sum_{i=2}^{n+1}H\left( X_{i},\ldots
,X_{n+1}|Y,X_{i-1}\right) +H\left( X_{1}|Y\right) =\sum_{i=1}^{n+1}H\left(
X_{i},\ldots ,X_{n+1}|Y,X_{i-1}\right) ,
\end{equation*}%
q.e.d.

\item The concavity of Shannon entropy, will be deduced by the concavity of
von Neumann's entropy in \ref{concavity_of_Shn_entr_by_VonNeum}. However
here is going to be demonstrated that binary entropy ($H_{\text{bin}}\left(
p\right) \triangleq H\left( p,1-p\right) $) is strictly concave,%
\begin{equation*}
H_{\text{bin}}\left( px_{1}+\left( 1-p\right) x_{2}\right) \geq pH_{\text{bin%
}}\left( x_{1}\right) +\left( 1-p\right) H_{\text{bin}}\left( x_{2}\right) ,
\end{equation*}%
where $0\leq p,x_{1},x_{2}\leq 1$ and equality holds for the trivial cases $%
x_{1}=x_{2},$ or $p=0,$ or $p=1.$ This is easily proved by using the fact
that the logarithmic function is increasing and $-p\left( 1-x\right) \geq
-\left( 1-px\right) ,$ hence%
\begin{eqnarray*}
H_{\text{bin}}\left( px_{1}+\left( 1-p\right) x_{2}\right) &\triangleq
&-\left( px_{1}+\left( 1-p\right) x_{2}\right) \log \left( px_{1}+\left(
1-p\right) x_{2}\right) \\
&&-\left[ 1-\left( px_{1}+\left( 1-p\right) x_{2}\right) \right] \log \left[
1-\left( px_{1}+\left( 1-p\right) x_{2}\right) \right] \\
&\geq &-px_{1}\log x_{1}-\left( 1-p\right) x_{2}\log x_{2}-p\left(
1-x_{1}\right) \log \left( 1-px_{1}\right) \\
&&-\left( 1-p\right) \left( 1-x_{2}\right) \log \left[ 1-\left( 1-p\right)
x_{2}\right] \\
&=&pH_{\text{bin}}\left( x_{1}\right) +\left( 1-p\right) H_{\text{bin}%
}\left( x_{2}\right) .
\end{eqnarray*}%
The strictness of concave property is seen by noting that only in the
trivial cases inequalities such as $\log px_{1}\leq \log \left(
px_{1}+\left( 1-p\right) x_{2}\right) $ could be equalities. Finally
concerning the binary entropy it is obvious that because $\frac{d}{dp}H_{%
\text{bin}}\left( p\right) =-\log p-1+\log \left( 1-p\right)
+1=0\Leftrightarrow p=\frac{1}{2},$ and for $p\neq 0,1,$ $\frac{d^{2}}{dp^{2}%
}H_{\text{bin}}\left( p\right) =-\frac{1}{p}-\frac{1}{1-p}<0,$ the maximum
is reached at $p=\frac{1}{2}.$
\end{enumerate}

\subsubsection{Some additional notes on the properties of Shannon entropy}

Concluding the properties of Shannon entropy, it should be noted that the
mutual information is not always subadditive or superadditive. One
counterexample for the first is the case where $X$ and $Y$ are independent
identically distributed random variables taking the values 0 or 1 with half
probability. Let $Z=X\oplus Y,$ where $\oplus $ the modulo 2 addition, then $%
H\left( X,Y:Z\right) =1$ and further calculating $H\left( X:Z\right)
+H\left( Y:Z\right) =0,$ that is%
\begin{equation*}
H\left( X,Y:Z\right) \nleq H\left( X:Z\right) +H\left( Y:Z\right) .
\end{equation*}%
The counterexample concerning the second case is the case of a random
variable $X_{1}$ taking values 0 or 1 with half probabilities and $%
X_{2}\triangleq Y_{1}\triangleq Y_{2}\triangleq X_{1}.$ Then $H\left(
X_{1}:Y_{1}\right) +H\left( X_{2}:Y_{2}\right) =2$ and in addition to this $%
H\left( X_{1},X_{2}:Y_{1},Y_{2}\right) =1,$ which means that%
\begin{equation*}
H\left( X_{1}:Y_{1}\right) +H\left( X_{2}:Y_{2}\right) \nleq H\left(
X_{1},X_{2}:Y_{1},Y_{2}\right) .
\end{equation*}

\section{Quantum physics and information theory\label%
{sec_Quant_phys_and_info_theory}}

Quantum theory is another very important area of physics, which is used to
describe the elementary particles that make up our world. The laws and the
intuition of quantum theory are totally different from the classical case.
To be more specific quantum theory is considered as counter intuitive, or
quoting Richard Feynman, \emph{''nobody really understands quantum
mechanics''}. However quantum physics offers new phenomena and properties
which can change peoples view for information. These properties are going to
be investigated in this section.

\subsection{Basic features of quantum mechanics relevant to information
theory\label{subsec_Basic_feat_of_quant_mech}}

Mathematically quantum entities are represented by Hilbert space vectors $%
\left| \psi \right\rangle ,$ usually normalized $\left\langle \psi |\psi
\right\rangle =1.$ Quantum systems evolve unitarily, that is, if a system is
initially in a state $\left| \psi _{1}\right\rangle ,$ it becomes later
another state $\left| \psi _{2}\right\rangle $ after a unitary operation%
\begin{equation}
U\left| \psi _{1}\right\rangle =\left| \psi _{2}\right\rangle .
\label{wave_fun_unitary_evolution}
\end{equation}%
\emph{Unitary operations are reversible}, since $UU^{\dagger }=1$ and
previous states can be reconstructed by $\left| \psi _{1}\right\rangle
=U^{\dagger }\left| \psi _{2}\right\rangle .$ What is important about such
operations is that because $\left\langle \psi _{2}|\psi _{2}\right\rangle
=\left\langle \psi _{1}\right| U^{\dagger }U\left| \psi _{1}\right\rangle
=\left\langle \psi _{1}\right| I\left| \psi _{1}\right\rangle =1,$
normalization, which soon will be interpreted as probability, is conserved.
The measurement of the properties of these objects is described by a
collection of operators $\left\{ M_{m}\right\} .$ The quantum object will
found to be in the $m$-th state with probability 
\begin{equation}
p\left( m\right) =\left\langle \psi \right| M_{m}^{\dagger }M_{m}\left| \psi
\right\rangle ,  \label{probability_of_measuring_Mm}
\end{equation}%
and after this measurement it is going to be in definite state, possibly
different from the starting one%
\begin{equation}
\frac{M_{m}}{\sqrt{\left\langle \psi \right| M_{m}^{\dagger }M_{m}\left|
\psi \right\rangle }}\left| \psi \right\rangle .
\label{state_after_measurement}
\end{equation}%
Of course $m$ should be viewed as the measurement outcome, hence information
extracted from a physical system. This way information content can be
assigned to each state of a quantum object. As a practical example, in each
energy state of an atom one can map four numbers or in each polarization
state of a photon one can map two numbers say 0 and 1. The last case is
similar to bits of classical information theory, but because a photon is a
quantum entity they are named qubits (quantum bits). It should be stressed
here that the measurement operators should satisfy the completeness relation 
$\sum_{m}M_{m}^{\dagger }M_{m}=I$ which results $\sum_{m}p\left( m\right)
=\sum_{m}\left\langle \psi \right| M_{m}^{\dagger }M_{m}\left| \psi
\right\rangle =1$ as instructed by probability theory. However this implies
that $\left\langle \psi \right| M_{m}^{\dagger }M_{m}\left| \psi
\right\rangle \neq 1$ and looking at equation (\ref{state_after_measurement}%
) one understands that \emph{measurement is an irreversible operation}.

What is very interesting about quantum entities is that they can either be
in a definite state or in a superposition of states! Mathematically this is
written%
\begin{equation*}
\left| \psi \right\rangle =\underset{s}{\sum }c_{s}\left| s\right\rangle .
\end{equation*}%
Using the language of quantum theory $\left| \psi \right\rangle $ is in
state $\left| s\right\rangle $ with probability $c_{s}^{\ast }c_{s},$ and
because of normalization the total probability of measuring $\left| \psi
\right\rangle $ is 
\begin{equation}
\left\langle \psi |\psi \right\rangle =\sum_{s}c_{s}^{\ast }c_{s}=1.
\label{wave_fun_normalization}
\end{equation}%
States $\left| s\right\rangle $ are usually orthonormal to each other, hence%
\begin{equation}
\left\langle \psi |s\right\rangle =c_{s}.
\label{wave_fun_coef_found_by_inner_product}
\end{equation}%
Although being simultaneously in many states sounds weird, quantum
information can be very powerful in computing. Suppose some quantum computer
takes as input quantum objects, which are in a superposition of multiple
states, then the output is going to be quantum objects which of course are
going to be in multiple states too. This way one can have many calculations
done only by one computational step! However careful extraction of results
is needed \cite{Deutsch,Shor94,Shor97,Grover96,Grover97}, because quantum
measurement has as outcome only one answer from the superposition of
multiple states, as equation (\ref{state_after_measurement}) instructs, and
further information is lost. Then one can have incredible results, like for
example calculate discrete logarithms and factorize numbers in polynomial
time \cite{Shor94,Shor97}, or search an unsorted database of $N$ objects
with only $\sqrt{N}$ iterations \cite{Grover96,Grover97}!

In contrast to classical information which under perfect conditions can be
known up to a desired accuracy, quantum information is sometimes ambiguous.
This is because one cannot distinguish non-orthogonal states reliably.
Assuming for a while that such a distinction is possible for two
non-orthogonal states $\left| \psi _{1}\right\rangle $ and $\left| \psi
_{2}\right\rangle $ and a collection of measurement operators $\left\{
M_{m}\right\} .$ Then according to this assumption some of the measuring
operators give reliable information whether the measured quantity is $\left|
\psi _{1}\right\rangle $ or $\left| \psi _{2}\right\rangle ,$ and collecting
them together the following two distinguishing POVM elements can be defined%
\begin{equation*}
E_{i}\triangleq \underset{M_{m}\text{ measuring }i}{\sum }M_{m}^{\dagger
}M_{m},\text{ }i=1,2.
\end{equation*}%
The assumption that these states can be reliably distinguished, is expressed
mathematically%
\begin{equation}
\left\langle \psi _{i}\right| E_{i}\left| \psi _{i}\right\rangle =1,\text{ }%
i=1,2.  \label{<psi_i|E_i|psi_i>}
\end{equation}%
Since $\sum_{i=1,2}E_{i}=I$ it follows that $\sum_{i=1,2}\left\langle \psi
_{1}\right| E_{i}\left| \psi _{1}\right\rangle =1.$ Because $E_{1}$ operator
reliable measures the first state, then $\left\langle \psi _{1}\right|
E_{1}\left| \psi _{1}\right\rangle =1,$ hence the other term must be 
\begin{equation}
\left\langle \psi _{1}\right| E_{2}\left| \psi _{1}\right\rangle =0.
\label{<psi_1|E_2|psi_1>=0}
\end{equation}%
Suppose $\left| \psi _{2}\right\rangle $ is decomposed in $\left| \psi
_{1}\right\rangle $ and and orthogonal state to $\left| \psi
_{1}\right\rangle ,$ say $\left| \psi _{\perp }\right\rangle ;$ then $\left|
\psi _{2}\right\rangle =\alpha \left| \psi _{1}\right\rangle +\beta \left|
\psi _{\perp }\right\rangle .$ Of course $\left| \alpha \right| ^{2}+\left|
\beta \right| ^{2}=1$ and $\left| \beta \right| <1$ because $\left| \psi
_{1}\right\rangle $ and $\left| \psi _{2}\right\rangle $ are not orthogonal.
Using the last decomposition and (\ref{<psi_1|E_2|psi_1>=0}) $\left\langle
\psi _{2}\right| E_{2}\left| \psi _{2}\right\rangle =\left| \beta \right|
^{2}\left\langle \psi _{\perp }\right| E_{2}\left| \psi _{\perp
}\right\rangle \leq \left| \beta \right| ^{2}<1$ which is in contradiction
with (\ref{<psi_i|E_i|psi_i>}).

In what concerns the results of the last paragraph it should be additionally
mentioned that information gain implies disturbance. Let $\left| \psi
\right\rangle $ and $\left| \phi \right\rangle $ be non-orthogonal states,
without loss of generality assume that a unitary process is used to obtain
information with the aid of an ancilla system $\left| u\right\rangle $.
Assuming that such a process does not disturb the system, then in both
cases, one obtains%
\begin{eqnarray*}
\left| \psi \right\rangle \otimes \left| u\right\rangle &\rightarrow &\left|
\psi \right\rangle \otimes \left| v\right\rangle , \\
\left| \phi \right\rangle \otimes \left| u\right\rangle &\rightarrow &\left|
\phi \right\rangle \otimes \left| v^{\prime }\right\rangle .
\end{eqnarray*}%
Then one would like $\left| v\right\rangle $ and $\left| v^{\prime
}\right\rangle $ to be different, in order to acquire information about the
states. However since the inner products are preserved under unitary
transformations, $\left\langle v|v^{\prime }\right\rangle \left\langle \psi
|\phi \right\rangle =\left\langle u|u\right\rangle \left\langle \psi |\phi
\right\rangle \Rightarrow \left\langle v|v^{\prime }\right\rangle
=\left\langle u|u\right\rangle =1,$ and hence are identical. Thus
distinguishing between $\left| \psi \right\rangle $ and $\left| \phi
\right\rangle $ must inevitably disturb at least one of these states.

However at least theoretically there is always a way of distinguishing
orthonormal states. Suppose $\left| i\right\rangle $ are orthonormal, then
it is straightforward to define the set of operators $M_{i}\triangleq \left|
i\right\rangle \left\langle i\right| $ plus the operator $M_{0}\triangleq 
\sqrt{I-\sum_{i\neq 0}\left| i\right\rangle \left\langle i\right| },$ which
satisfy the completeness relation. Now if the state $\left| i\right\rangle $
is prepared then $p\left( i\right) =\left\langle i\right| M_{i}^{\dagger
}M_{i}\left| i\right\rangle =1,$ thus they are reliably distinguished.

Another very surprising result is the prohibition of copying arbitrary
quantum states. This is known as no-cloning theorem \cite{Dieks,WootersZurek}
and it can be very easily proven. Suppose it is possible to have a quantum
photocopying machine, which will have as input a quantum white paper $\left|
w\right\rangle $ and a state to be copied. The quantum photocopying machine
should be realized by a unitary transformation $U$ and if somebody tries to
photocopy two states $\left| \psi \right\rangle $ and $\left| \phi
\right\rangle ,$ it should very naturally work as follows%
\begin{eqnarray*}
U\left\{ \left| w\right\rangle \otimes \left| \psi \right\rangle \right\}
&=&\left| \psi \right\rangle \otimes \left| \psi \right\rangle , \\
U\left\{ \left| w\right\rangle \otimes \left| \phi \right\rangle \right\}
&=&\left| \phi \right\rangle \otimes \left| \phi \right\rangle .
\end{eqnarray*}%
Now taking the inner product of these relations $\left\langle \psi |\phi
\right\rangle =\left\langle \psi |\phi \right\rangle ^{2},$ thus $%
\left\langle \psi |\phi \right\rangle =0$ or $\left\langle \psi |\phi
\right\rangle =1,$ hence $\left| \psi \right\rangle =\left| \phi
\right\rangle $ or $\left| \psi \right\rangle $ and $\left| \phi
\right\rangle $ are orthogonal. This means that cloning is allowed only for
orthogonal states! Thus at least somebody can construct a device, by quantum
circuits, to copy orthogonal states. For example if $\left| \psi
\right\rangle $ and $\left| \phi \right\rangle $ are orthogonal then there
exists a unitary transformation $U$ such that $U\left| \psi \right\rangle
=\left| 0\right\rangle $ and $U\left| \phi \right\rangle =\left|
1\right\rangle .$ Then by applying the \texttt{FANOUT} quantum gate, which
maps the input to $\left| 00\right\rangle $ if the input qubit was $\left|
0\right\rangle $ and to $\left| 11\right\rangle $ if the it was $\left|
1\right\rangle ,$ and further applying $U^{\dagger }\otimes U^{\dagger }$ to
the tensor product of outcoming qubits, then either the state $\left| \psi
\right\rangle $ is will be copied and finally get $\left| \psi \psi
\right\rangle ,$ or $\left| \phi \right\rangle $ and get $\left| \phi \phi
\right\rangle .$

The basic information relevant features of quantum mechanics, analyzed in
this subsection, are summarized in Figure \ref{fig_Features_of_Quant_Mech}.

\begin{figure}[tbh]
\begin{center}
\fbox{%
\begin{tabular}{c}
{\LARGE Basic features of quantum mechanics} \\ \hline
\\ 
\multicolumn{1}{l}{\large 1. Reversibility of quantum operations} \\ 
\multicolumn{1}{l}{\large 2. Irreversibility of measurement} \\ 
\multicolumn{1}{l}{\large 3. Probabilistic outcomes} \\ 
\multicolumn{1}{l}{\large 4. Superposition of states} \\ 
\multicolumn{1}{l}{\large 5. Distinguishability of orthogonal states by
operators} \\ 
\multicolumn{1}{l}{\large 6. Non-distinguishability of non-orthogonal states}
\\ 
\multicolumn{1}{l}{\large 7. Information gain implies disturbance} \\ 
\multicolumn{1}{l}{\large 8. Non-cloning of non-orthogonal states}%
\end{tabular}%
}
\end{center}
\caption{A summary of basic information relative features, of quantum
mechanics.}
\label{fig_Features_of_Quant_Mech}
\end{figure}

\subsection{The language of quantum information theory: quantum
thermodynamics\label{subsec_Quantum_thermodynamics}}

As in the case of classical information theory thermodynamics should be
studied for abstracting quantum information. As it was already stated in the
last paragraphs, quantum mechanics have some sort of built-in probabilities.
However this is not enough. The probabilistic features of quantum states are
different from that of classical thermodynamics. The difference is
demonstrated by taking two totally independent facts $A$ and $B.$ Then the
probability of both of them occurring would be $P_{AB}^{\text{classical}%
}=P\left( A\right) +P\left( B\right) .$ In contrast, in quantum mechanics
the wave functions should be added, and by calculating the inner product
probabilities are obtained. More analytically%
\begin{eqnarray}
P_{AB}^{\text{quantum}} &=&\left( \left\langle \psi _{A}\right|
+\left\langle \psi _{B}\right| \right) \left( \left| \psi _{A}\right\rangle
+\left| \psi _{B}\right\rangle \right) =\left( \left\langle \psi _{A}|\psi
_{A}\right\rangle +\left\langle \psi _{B}|\psi _{B}\right\rangle \right) +2%
\text{Re}\left\langle \psi _{A}^{\ast }|\psi _{B}\right\rangle
\label{difference_of_classical&quant_probab} \\
&=&P\left( A\right) +P\left( B\right) +2\text{Re}\left\langle \psi
_{A}^{\ast }|\psi _{B}\right\rangle \overset{\text{in general}}{\neq }%
P_{AB}^{\text{classical}}.  \notag
\end{eqnarray}%
Moreover if somebody decides to encode some information with quantum objects
then he is going to be interested with probabilities of occurrence of the
alphabet, exactly the same way it was done in section \ref%
{sec_Classical_physics_&_info_th}, and he would never like to mess up with
the probabilities already found in quantum entities. For this reason a
quantum version of thermodynamics is needed. In the sequel the name
thermodynamical probabilities is used to distinguish the statistical mixture
of several quantum states, from the quantum probabilities occurring by
observation of a quantum state.

The basic tool of quantum thermodynamics is the density matrix and its
simplest case is when the quantum state occurs with thermodynamical
probability $p=1.$ Then by (\ref{wave_fun_coef_found_by_inner_product}) $%
\left\langle t|\psi \right\rangle \left\langle \psi |s\right\rangle
=c_{t}^{\ast }c_{s}$ and therefore%
\begin{equation*}
\rho \triangleq \left| \psi \right\rangle \left\langle \psi \right| ,
\end{equation*}%
is the natural matrix generalization of a quantum vector state. This was the
definition of a density matrix of a \emph{pure} state, in contradiction to 
\emph{mixture} of states where several states occur with probabilities $%
0<p<1.$ Each element of this matrix is%
\begin{equation*}
\rho _{ts}\triangleq \left\langle t|\psi \right\rangle \left\langle \psi
|s\right\rangle =c_{t}^{\ast }c_{s}.
\end{equation*}%
and by equation (\ref{wave_fun_normalization}) where the normalization of
vector was defined, density matrix is correspondingly normalized by%
\begin{equation}
\text{tr}\rho =\sum_{s}\rho _{ss}=\sum_{s}c_{s}^{\ast
}c_{s}=\sum_{s}\left\langle s|\psi \right\rangle \left\langle \psi
|s\right\rangle =1\text{.}  \label{dens_matr_norm}
\end{equation}%
Moreover it is straightforward that unitary evolution (\ref%
{wave_fun_unitary_evolution}) is described by%
\begin{equation}
U\left| \psi \right\rangle \left\langle \psi \right| U^{\dagger }=U\rho
U^{\dagger },  \label{dens_matr_unitary_evol}
\end{equation}%
the probability for measuring the $m$-th state (\ref%
{probability_of_measuring_Mm}) is given by%
\begin{equation}
p\left( m\right) =\left\langle \psi \right| M_{m}^{\dagger }M_{m}\left| \psi
\right\rangle =\left\langle \psi \right| M_{m}^{\dagger }\sum_{s}\left|
s\right\rangle \left\langle s\right| M_{m}\left| \psi \right\rangle
=\sum_{s}\left\langle \psi \right| M_{m}^{\dagger }\left| s\right\rangle
\left\langle s\right| M_{m}\left| \psi \right\rangle =\text{tr}\left(
M_{m}^{\dagger }\rho M_{m}\right) ,  \label{dens_matr_prob_of_measur}
\end{equation}%
and the state after measurement (\ref{state_after_measurement}) is obtained
by%
\begin{equation}
\frac{M_{m}}{\sqrt{\left\langle \psi \right| M_{m}^{\dagger }M_{m}\left|
\psi \right\rangle }}\left| \psi \right\rangle \left\langle \psi \right| 
\frac{M_{m}^{\dagger }}{\sqrt{\left\langle \psi \right| M_{m}^{\dagger
}M_{m}\left| \psi \right\rangle }}=\frac{1}{\text{tr}\left( M_{m}^{\dagger
}\rho M_{m}\right) }M_{m}\rho M_{m}^{\dagger }.  \label{dens_matr_measur}
\end{equation}

Suppose now there is a collection of quantum states $\left| \psi
_{i}\right\rangle $ occurring with thermodynamical probabilities $p_{i},$
then the mixed density operator is simply defined%
\begin{equation}
\rho \triangleq \sum_{i}p_{i}\left| \psi _{i}\right\rangle \left\langle \psi
_{i}\right| .  \label{def_density_matrix}
\end{equation}%
This construction is precisely what was expected in the beginning of this
subsection, because the thermodynamical probabilities are real numbers and
can be added, contrasting to the situation of quantum probabilities for a
state vector (\ref{difference_of_classical&quant_probab}). The
generalization of normalization (\ref{dens_matr_norm}), unitary evolution (%
\ref{dens_matr_unitary_evol}), probability of measurement (\ref%
{dens_matr_prob_of_measur}) and measurement (\ref{dens_matr_measur}) for the
mixed density matrix are%
\begin{gather*}
\text{tr}\rho =\sum_{i}p_{i}\text{tr}\left( \left| \psi _{i}\right\rangle
\left\langle \psi _{i}\right| \right) =1, \\
\sum_{i}p_{i}\left| \psi _{i}\right\rangle \left\langle \psi _{i}\right|
=U\rho U^{\dagger }, \\
p\left( m\right) =\text{tr}\left( M_{m}^{\dagger }\rho M_{m}\right) , \\
\frac{1}{\text{tr}\left( M_{m}^{\dagger }\rho M_{m}\right) }M_{m}\rho
M_{m}^{\dagger }.
\end{gather*}%
The above results are pretty obvious for the first, the second and the
fourth, but some algebra of probabilities is needed for the third (refer to %
\cite[p.99]{NielsenChuang} for details). The unitary evolution of the system
can be generalized by \emph{quantum operations}%
\begin{equation*}
\mathcal{E}\left( \rho \right) =\sum_{i}E_{i}^{\dagger }\rho E_{i},
\end{equation*}%
where $\sum_{i}E_{i}^{\dagger }E_{i}=I.$ Quantum operations are sometimes
referred in the literature \cite{Preskill} as \emph{superoperators}.

Quantum thermodynamics, just described can be viewed as a transition from
classical to quantum thermodynamics. Suppose an ensemble of $n$ particles in
an equilibrium is given. For this ensemble assume that the $i$-th particle
is located at $\mathbf{x}_{i},$ has velocity $\mathbf{v}_{i}$ and energy $%
E_{i}\left( \mathbf{x}_{i},\mathbf{v}_{i}\right) .$ Classical thermodynamics
says that for the $i$-th particle, there is a probability 
\begin{equation}
p_{i}=\frac{1}{\sum_{j}e^{-\beta E_{j}}}e^{-\beta E_{i}},
\label{thermodynamical_probability}
\end{equation}%
to be in this state, where $\beta =\frac{1}{k_{B}T},$ with $k_{B}$
Boltzmann's constant and $T$ the temperature. Now if the particles where
described by quantum mechanics, then if the $i$-th would have an eigenenergy 
$E_{i},$ given by the solution of the Schr\"{o}dinger equation $H\left| \psi
_{i}\right\rangle =E_{i}\left| \psi _{i}\right\rangle ,$ where $H$ is the
Hamiltonian of the system. Now with the help of equation (\ref%
{thermodynamical_probability}) the density matrix as was defined in (\ref%
{def_density_matrix}) can be written as%
\begin{equation*}
\rho =\sum_{i}\left| \psi _{i}\right\rangle p_{i}\left\langle \psi
_{i}\right| =\frac{1}{\sum_{j}e^{-\beta E_{j}}}\sum_{i}\left| \psi
_{i}\right\rangle e^{-\beta E_{i}}\left\langle \psi _{i}\right| =\frac{1}{%
\text{tr}e^{-\beta H}}e^{-\beta H},
\end{equation*}%
which is a generalization of classical thermodynamical probability (\ref%
{thermodynamical_probability}). This transition is illustrated in Figure \ref%
{fig_From_Class_to_Quant_Thermodynamics}.

\begin{figure}[tbh]
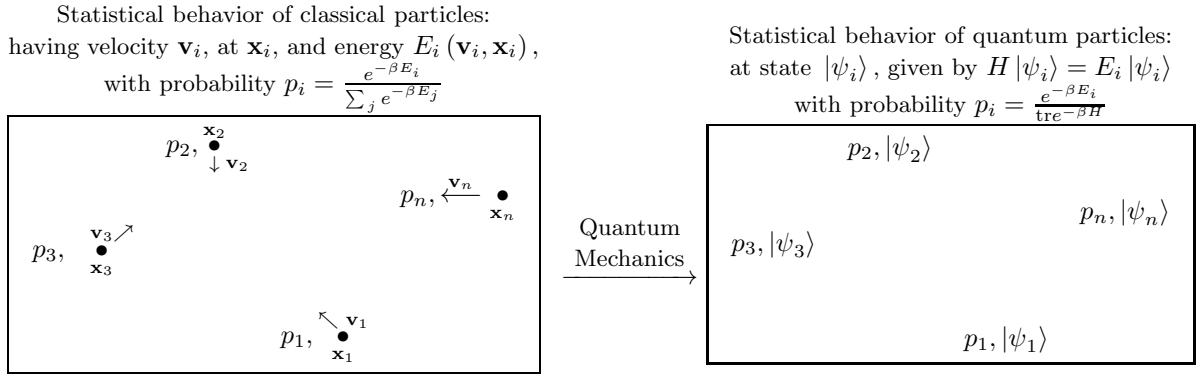

\begin{center}
\begin{tabular}{c}
$\overset{%
\begin{array}{c}
\text{{\small Statistical behavior of classical particles:}} \\ 
\text{{\small having velocity }}\mathbf{v}_{i}\text{{\small , at }}\mathbf{x}%
_{i}\text{{\small , and energy }}E_{i}\left( \mathbf{v}_{i},\mathbf{x}%
_{i}\right) , \\ 
\text{{\small with probability }}p_{i}=\frac{e^{-\beta E_{i}}}{%
\sum_{j}e^{-\beta E_{j}}}%
\end{array}%
}{\fbox{%
\begin{tabular}{cccc}
& $p_{2},\underset{\downarrow }{\overset{\mathbf{x}_{2}}{\bullet }}\underset{%
\mathbf{v}_{2}}{}$ &  &  \\ 
&  &  & $p_{n},\overset{\mathbf{v}_{n}}{\longleftarrow }\underset{\mathbf{x}%
_{n}}{\bullet }$ \\ 
$p_{3},\underset{\mathbf{x}_{3}}{\overset{\text{\quad }\mathbf{v}%
_{3}\nearrow }{\bullet }}$ &  &  &  \\ 
&  &  &  \\ 
&  & $p_{1},\overset{\nwarrow \text{ }\mathbf{v}_{1}}{\underset{\mathbf{x}%
_{1}}{\bullet }}$ & 
\end{tabular}%
}}\ \underrightarrow{%
\begin{array}{c}
\text{{\small Quantum}} \\ 
\text{{\small Mechanics}}%
\end{array}%
}$\ $\overset{%
\begin{array}{c}
\text{{\small Statistical behavior of quantum particles:}} \\ 
\text{{\small at state }}\left| \psi _{i}\right\rangle \text{{\small , given
by} }H\left| \psi _{i}\right\rangle =E_{i}\left| \psi _{i}\right\rangle \\ 
\text{{\small with probability }}p_{i}=\frac{e^{-\beta E_{i}}}{\text{tr}%
e^{-\beta H}}%
\end{array}%
}{\fbox{%
\begin{tabular}{cccc}
& $p_{2},\left| \psi _{2}\right\rangle $ &  &  \\ 
&  &  &  \\ 
&  &  & $p_{n},\left| \psi _{n}\right\rangle $ \\ 
$p_{3},\left| \psi _{3}\right\rangle $ &  &  &  \\ 
&  &  &  \\ 
&  &  &  \\ 
&  & $p_{1},\left| \psi _{1}\right\rangle $ & 
\end{tabular}%
}}$%
\end{tabular}%
\end{center}
\caption{Transition from classical to quantum thermodynamics.}
\label{fig_From_Class_to_Quant_Thermodynamics}
\end{figure}

\subsection{Von Neumann entropy and relevant measures of quantum information
theory}

Willing to describe quantum information, one could use quantum version of
entropy, and in order to justify its mathematical definition, recall how
Shannon entropy was given in equation (\ref{def_entropy}), and assume that a
density matrix $\rho $ is diagonalized by $\rho =\sum_{x}\lambda _{x}\left|
x\right\rangle \left\langle x\right| .$ Then naturally quantum entropy of $%
\rho $ is defined%
\begin{equation}
S\left( \rho \right) \triangleq -\underset{x}{\sum }\lambda _{x}\log \lambda
_{x}.  \label{def_VonNeym_entropy_by_eigenvalues}
\end{equation}%
Translating this into the mathematical formalism developed in the last
subsection, the \emph{von Neumann entropy} is defined%
\begin{equation}
S\left( \rho \right) \triangleq -\text{tr}\left( \rho \log \rho \right) .
\label{def_VonNeumann_entropy}
\end{equation}%
The last formula is often used for proving theoretical results and equation (%
\ref{def_VonNeym_entropy_by_eigenvalues}) is used for calculations. As an
example the von Neumann entropy of $\rho =p\left| 0\right\rangle
\left\langle 0\right| +\left( 1-p\right) \frac{\left( \left| 0\right\rangle
+\left| 1\right\rangle \right) \left( \left\langle 0\right| +\left\langle
1\right| \right) }{2}$ is found to be%
\begin{equation*}
S\left( \rho \right) =S\left( \left[ 
\begin{array}{cc}
p+\frac{1-p}{2} & \frac{1-p}{2} \\ 
\frac{1-p}{2} & \frac{1-p}{2}%
\end{array}%
\right] \right) =-\lambda _{1}\log \lambda _{1}-\lambda _{2}\log \lambda
_{2},
\end{equation*}%
where $\lambda _{1}=\frac{1+\sqrt{\left( 1+2p^{2}-2p\right) }}{2}$ and $%
\lambda _{2}=\frac{1-\sqrt{\left( 1+2p^{2}-2p\right) }}{2}$ are the
eigenvalues of the corresponding matrix. Surprisingly $S\left( \rho \right)
\neq H\left( p,1-p\right) $ even if the same probabilities where assigned
for both of them! This shows that quantum probabilities are not expelled by
quantum thermodynamics. The equality could only hold if the probabilities
written in Shannon's entropy are the eigenvalues of the density matrix.

Following the same path as for classical information theory, in the quantum
case it is straightforward to define the \emph{joint entropy}, the \emph{%
relative entropy} of $\rho $ to $\sigma ,$ the \emph{entropy of A
conditional on knowing B} and the \emph{common} or \emph{mutual information
of A and B}. Each case is correspondingly 
\begin{subequations}
\label{def_VonNeum_entorpies_4_case}
\begin{eqnarray}
S\left( A,B\right) &\triangleq &\text{tr}\left( \rho ^{AB}\log \rho
^{AB}\right) ,  \label{def_VonNeum_joint} \\
S\left( \rho ||\sigma \right) &\triangleq &\text{tr}\left( \rho \log \rho
\right) -\text{tr}\left( \rho \log \sigma \right) ,
\label{def_VonNeum_relative} \\
S\left( A|B\right) &\triangleq &S\left( A,B\right) -S\left( B\right) ,
\label{def_VonNeum_conditional} \\
S\left( A:B\right) &\triangleq &S\left( A\right) +S\left( B\right) -S\left(
A,B\right) =S\left( A\right) -S\left( A|B\right) .
\label{def_VonNeum_mutual}
\end{eqnarray}%
One can see that there are lot of similarities between Shannon's and von
Neumann's entropy. As such one can prove a result reminding equation (\ref%
{rel_entropy_metric_like_properties}) 
\end{subequations}
\begin{subequations}
\begin{eqnarray}
S\left( \rho ||\sigma \right) &>&0,  \label{Klein_inequality(inequality)} \\
S\left( \rho ||\sigma \right) &=&0\Leftrightarrow \rho =\sigma ,
\label{Klein_inequality(equality)}
\end{eqnarray}%
and is known as \emph{Klein's inequality}. This inequality provides evidence
of why von Neumann relative entropy is close to the notion of metric. What
is also important is that it can be used, like in the classic case, to prove
something corresponding to equation (\ref{entopy<_relation}) of Shannon's
entropy 
\end{subequations}
\begin{subequations}
\label{S(rho)<=logd}
\begin{eqnarray}
S\left( \rho \right) &<&\log d,  \label{S(rho)<=logd(<)} \\
S(\rho ) &=&\log d\Leftrightarrow \rho =\frac{1}{d}I.
\label{S(rho)<=logd(=)}
\end{eqnarray}%
In addition to this from the definition (\ref{def_VonNeumann_entropy}) it
follows that 
\end{subequations}
\begin{subequations}
\label{S(rho)>=0}
\begin{eqnarray}
S\left( \rho \right) &>&0,  \label{S(rho)>0} \\
S\left( \rho \right) &=&0\Leftrightarrow \rho \text{ is pure,}
\label{S(rho)=0}
\end{eqnarray}%
which resembles to equation (\ref{entropy>_relation}).

One can also prove that supposing some $\rho _{i}$ states, with
probabilities $p_{i},$ have support on orthogonal subspaces, then 
\end{subequations}
\begin{equation*}
S\left( \underset{i}{\sum }p_{i}\rho _{i}\right) =H\left( p_{i}\right) +%
\underset{i}{\sum }p_{i}S\left( \rho _{i}\right) .
\end{equation*}%
Directly from this relation the \textbf{joint entropy theorem} can be
proven, where supposing that $p_{i}$ are probabilities, $\left|
i\right\rangle $ are orthogonal states for a system $A$, and $\rho _{i}$ is
an set of density matrices of another system $B$, then 
\begin{equation}
S\left( \underset{i}{\sum }p_{i}\left| i\right\rangle \left\langle i\right|
\otimes \rho _{i}\right) =H\left( p_{i}\right) +\underset{i}{\sum }%
p_{i}S\left( \rho _{i}\right) .  \label{Joint_entropy_theorem}
\end{equation}%
Using the definition of von Neumann entropy (\ref{def_VonNeumann_entropy})
and the above mentioned theorem for the case where $\rho _{i}=\sigma $ for
every $i,$ and let $\rho $ be a density matrix with eigenvalues $p_{i},$ and
eigenvectors $\left| i\right\rangle ,$ then the entropy of a tensor product $%
\rho \otimes \sigma $ is found to be%
\begin{equation}
S\left( \rho \otimes \sigma \right) =S\left( \rho \right) +S\left( \sigma
\right) .  \label{entropy_of_tensor_product}
\end{equation}

Another interesting result can be derived by Schmidt decomposition; if a
composite system $AB$ is in a pure state, it has subsystems $A$ and $B$ with
density matrices of equal eigenvalues, and by (\ref%
{def_VonNeym_entropy_by_eigenvalues})%
\begin{equation}
S\left( A\right) =S\left( B\right) .  \label{pure_states_entropy}
\end{equation}

\subsection{A great discrepancy between classical and quantum information
theory: quantum entanglement\label{subsec_QuantumEntanglement}}

The tools developed in the above subsection can help reveal a great
discrepancy between classical and quantum information theory: \emph{%
entanglement}. Concerning the nomenclature, in quantum mechanics two states
are named entangled if they cannot be written as a tensor product of other
states. For the demonstration of the aforementioned discrepancy, let for
example a composite system $AB$ be in an entangled pure state $\left|
AB\right\rangle ,$ then, because of entanglement, in the Schmidt
decomposition it should be written as the sum of more than one terms 
\begin{equation}
\left| AB\right\rangle =\underset{i\in I}{\sum }\lambda _{i}\left|
i_{A}\right\rangle \left| i_{B}\right\rangle ,\text{ with }\left| I\right|
>1,  \label{|AB>_pure&entangled}
\end{equation}%
where $\left| i_{A}\right\rangle $ and $\left| i_{B}\right\rangle $ are
orthonormal bases. The corresponding density matrix is obviously $\rho
^{AB}=\left| AB\right\rangle \left\langle AB\right| =\sum_{i,j\in I}\lambda
_{i}\lambda _{j}\left| i_{A}\right\rangle \left| i_{B}\right\rangle
\left\langle j_{A}\right| \left\langle j_{B}\right| .$ As usually the
density matrix of the subsystem $B$ can be found by tracing out system $A,$ 
\begin{equation*}
\rho ^{B}=\text{tr}_{A}\left( \rho ^{AB}\right) =\underset{i,j,k\in I}{\sum }%
\lambda _{i}\lambda _{j}\left\langle k_{A}|i_{A}\right\rangle \left|
i_{B}\right\rangle \left\langle j_{A}|k_{A}\right\rangle \left\langle
j_{B}\right| =\underset{i\in I}{\sum }\lambda _{i}^{2}\left|
i_{B}\right\rangle \left\langle i_{B}\right| .
\end{equation*}%
Now because of the assumption $\left| I\right| >1$ in (\ref%
{|AB>_pure&entangled}) and the fact that $\left| i_{B}\right\rangle $ are
orthonormal bases and it is impossible to collect them together in a tensor
product, subsystem $B$ is not pure. Thus by equation (\ref{S(rho)>0}) $%
S\left( B\right) >0,$ $AB$ is pure thus by (\ref{S(rho)=0}) $S\left(
A,B\right) =0$ and obviously by (\ref{def_VonNeum_relative}) $S\left(
A|B\right) <0.$ The last steps can be repeated backwards and the conclusion
which can be drawn is that a pure composite system $AB$ is entangled if and
only if $S\left( A|B\right) <0.$

Of course in classical information theory conditional entropy could only be $%
H\left( X|Y\right) \geq 0$ (property 2 in subsection \ref%
{sec_BasicPropertiesofShannonEntropy}) and that is obviously the reason why
entangled states did not exist at all! This is an exclusive feature of
quantum information theory. A very intriguing or better to say a very
entangled feature, which until now is not well understood by physicists.
However it has incredible applications, such as quantum cryptography, which
will be the main topic of the chapter \ref{chap_Quantum_cryptography}.
Concerning the nomenclature, entangled states are named after the fact that $%
S\left( A|B\right) <0\overset{(\ref{def_VonNeum_conditional})}{%
\Longleftrightarrow }S\left( A,B\right) <S\left( B\right) $ which means that
the ignorance about a system $B$ can be in quantum mechanics more than the
ignorance of both $A$ and $B!$ This proposes some correlation between these
two systems.

How can nature have such a peculiar property? Imagine a simple pair of
quantum particles, with two possible states each $\left| 0\right\rangle $
and $\left| 1\right\rangle .$ Then a possible formulation of entanglement
can be a state $\left| \psi \right\rangle =\frac{\left| 0\right\rangle
\otimes \left| 0\right\rangle }{\sqrt{2}}+\frac{\left| 1\right\rangle
\otimes \left| 1\right\rangle }{\sqrt{2}}.$ After a measurement $M_{m}$ of
the first particle for example, according to (\ref{state_after_measurement})%
\begin{equation*}
M_{m}\left| \psi \right\rangle =1\left( \left| m\right\rangle \otimes \left|
m\right\rangle \right) +0\left( \left| 1-m\right\rangle \otimes \left|
1-m\right\rangle \right) =\left| m\right\rangle \otimes \left|
m\right\rangle ,
\end{equation*}%
hence they both collapse to state $\left| m\right\rangle .$ This example
sheds light into the quantum property, where ignorance of both particles is
greater than the ignorance of one of them, since perfect knowledge about one
implies perfect knowledge about the second.

\subsection{Basic properties of von Neumann entropy}

The basic properties of von Neumann entropy, which can be compared to the
properties of Shannon entropy discussed in subsection \ref%
{sec_BasicPropertiesofShannonEntropy}, are:

\begin{enumerate}
\item $S\left( A,B\right) =S\left( B,A\right) ,$ $S\left( A:B\right)
=S\left( B:A\right) .$

\item \textbf{Unitary operations preserve entropy:} $S\left( U\rho
U^{\dagger }\right) =S\left( \rho \right) .\label%
{Unitary_operation_preserve_entropy}$

\item \textbf{Subadditivity:} $S\left( A,B\right) \leq S\left( A\right)
+S\left( B\right) .$

\item $S\left( A,B\right) \geq \left| S\left( A\right) -S\left( B\right)
\right| $ (\emph{Triangle} or \emph{Araki-Lieb} inequality).

\item \textbf{Strict concavity of the entropy:} Suppose there are
probabilities $p_{i}\geq 0$ and the corresponding density matrices $\rho
_{i},$ then $S\left( \sum_{i}p_{i}\rho _{i}\right) <\sum_{i}p_{i}S\left(
\rho _{i}\right) $ and $S\left( \sum_{i}p_{i}\rho _{i}\right)
=\sum_{i}p_{i}S\left( \rho _{i}\right) \Leftrightarrow \rho _{i}$ for which $%
p_{i}>0$ are all identical.

\item \textbf{Upper bound of a mixture of states:} Suppose $\rho
=\sum_{i}p_{i}\rho _{i}$ where $p_{i}$ probabilities and the corresponding
density matrices $\rho _{i},$ then $S\left( \rho \right)
<\sum_{i}p_{i}S\left( \rho _{i}\right) +H\left( p_{i}\right) $ and $S\left(
\rho \right) =\sum_{i}p_{i}S\left( \rho _{i}\right) +H\left( p_{i}\right)
\Leftrightarrow \rho _{i}$ have support on orthogonal subspaces.

\item \textbf{Strong subadditivity:} $S\left( A,B,C\right) +S\left( B\right)
\leq S\left( A,B\right) +S\left( B,C\right) ,$ or equivalently $S\left(
A\right) +S\left( B\right) \leq S\left( A,C\right) +S\left( B,C\right) .$

\item \textbf{Conditioning reduces entropy:} $S\left( A|B,C\right) \leq
S\left( A|B\right) .$

\item \textbf{Discarding quantum systems never increases mutual information:}
Suppose $ABC$ is a composite quantum system, then $S\left( A:B\right) \leq
S\left( A:B,C\right) .$

\item \textbf{Trace preserving quantum operations never increase mutual
information:} Suppose $AB$ is a composite quantum system and $\mathcal{E}$
is a trace preserving quantum operation on system $B.$ Let $S\left(
A:B\right) $ denote the mutual information between systems $A$ and $\ B$
before $\mathcal{E}$ applied to system $\ B,$ and $S\left( A^{\prime
}:B^{\prime }\right) $ the mutual information after $\mathcal{E}$ is applied
to system $B.$ Then $S\left( A^{\prime }:B^{\prime }\right) \leq S\left(
A:B\right) .$

\item \textbf{Relative entropy is jointly convex in its arguments: }let $%
0\leq \lambda \leq 1,$ then 
\begin{equation*}
S\left( \lambda A_{1}+\left( 1-\lambda \right) A_{2}||\lambda B_{1}+\left(
1-\lambda \right) B_{2}\right) \geq \lambda S\left( A_{1}||B_{1}\right)
+\left( 1-\lambda \right) S\left( A_{2}||B_{2}\right) .
\end{equation*}

\item \textbf{The relative entropy is monotonic:} $S\left( \rho ^{A}||\sigma
^{A}\right) \leq S\left( \rho ^{AB}||\sigma ^{AB}\right) .$
\end{enumerate}

\emph{Proof}

\begin{enumerate}
\item Obvious from the definition of joint entropy (\ref{def_VonNeum_joint})
and mutual entropy (\ref{def_VonNeum_mutual}).

\item Let $U$ be a unitary matrix then $S\left( U\rho U^{\dagger }\right)
\triangleq -tr\left[ U\rho U^{\dagger }\log \left( U\rho U^{\dagger }\right) %
\right] =-tr\left[ U\rho U^{\dagger }U\left( \log \rho \right) U^{\dagger }%
\right] ,$ where the fact that $U\rho U^{\dagger }$ and $\rho $ are similar
and hence they have the same eigenvalues, was employed. Furthermore $U$ is
unitary, hence $UU^{\dagger }=I,$ and the proof is concluded $S\left( U\rho
U^{\dagger }\right) =-tr\left[ U\rho \left( \log \rho \right) U^{\dagger }%
\right] \overset{\text{tr}\left( AB\right) =\text{tr}\left( BA\right) }{=}-tr%
\left[ U^{\dagger }U\rho \left( \log \rho \right) \right] \overset{%
U^{\dagger }U=I}{=}-tr\left( \rho \log \rho \right) \triangleq S\left( \rho
\right) .$

\item Refer to \cite[p.516]{NielsenChuang}.

\item Assume a fictitious state $R$ purifying the system $\rho ^{AB},$ then
by applying subadditivity (property 3) $S\left( R\right) +S\left( A\right)
\geq S\left( A,R\right) .$ Because $\rho ^{ABR}$ is pure, and by (\ref%
{pure_states_entropy}) $S\left( R\right) =S\left( A,B\right) $ and $S\left(
A,R\right) =S\left( B\right) .$ Combining the last two equations with the
last inequality, $S\left( A,B\right) \geq S\left( B\right) -S\left( A\right)
.$ Moreover because $S\left( A,B\right) =S\left( B,A\right) ,$ $A$ and $B$
can be interchanged and then $S\left( A,B\right) \geq S\left( A\right)
-S\left( B\right) ,$ thus $S\left( A,B\right) \geq \left| S\left( A\right)
-S\left( B\right) \right| .$ It is obvious by (\ref%
{entropy_of_tensor_product}) that the equality holds if $\rho ^{AR}=\rho
^{A}\otimes \rho ^{B}.$ This is hard to understand because $R$ system was
artificially introduced. Another way to obtain the equality condition is by
assuming $\rho ^{AB}$ has a spectral decomposition $\rho
^{AB}=\sum_{ik}\lambda _{ik}\left| i_{A}\right\rangle \left\langle
i_{A}\right| \otimes \left| k_{B}\right\rangle \left\langle k_{B}\right| ,$
and obviously $\rho ^{A}=$tr$_{B}\rho ^{AB}=\sum_{i}\left( \sum_{k}\lambda
_{ik}\right) \left| i_{A}\right\rangle \left\langle i_{A}\right| $ and $\rho
^{B}=\sum_{k}\left( \sum_{i}\lambda _{ik}\right) \left| k_{B}\right\rangle
\left\langle k_{B}\right| ,$ then one can write%
\begin{equation*}
S\left( A,B\right) =S\left( B\right) -S\left( A\right) \Leftrightarrow
\sum_{ik}\lambda _{ik}\log \lambda _{ik}=\sum_{ik}\lambda _{ik}\log \left( 
\frac{\sum_{j}\lambda _{jk}}{\sum_{m}\lambda _{im}}\right) \Leftrightarrow
\lambda _{ik}=\frac{\sum_{j}\lambda _{jk}}{\sum_{m}\lambda _{im}}.
\end{equation*}%
Summing over $k$ in the last equation%
\begin{equation*}
\sum_{k}\lambda _{ik}=\frac{\sum_{kj}\lambda _{jk}}{\sum_{m}\lambda _{im}}=%
\frac{1}{\sum_{m}\lambda _{im}}\Leftrightarrow \left( \sum_{k}\lambda
_{ik}\right) ^{2}=1\Leftrightarrow \sum_{k}\lambda _{ik}=1,
\end{equation*}%
where the last relation holds because $\lambda _{ik}\geq 0.$ Now combining
the last two outcomes%
\begin{equation*}
S\left( A,B\right) =S\left( B\right) -S\left( A\right) \Leftrightarrow
\lambda _{ik}=\sum_{j}\lambda _{jk}\Leftrightarrow \lambda _{ik}\equiv
\lambda _{i}.
\end{equation*}%
Rephrasing this result the equality condition holds if and only if the
matrices tr$_{B}\left( \left| i\right\rangle \left\langle i\right| \right) $
have a common eigenbasis and the matrices tr$_{A}\left( \left|
i\right\rangle \left\langle i\right| \right) $ have orthogonal support. As
an example of the above comment consider the systems $\rho ^{A}=\left|
0\right\rangle \left\langle 0\right| $ and $\rho ^{B}=\frac{1}{2}\left|
0\right\rangle \left\langle 0\right| +\frac{1}{2}\left| 1\right\rangle
\left\langle 1\right| ,$ then the entropy of each is $S\left( A\right) =0$
and $S\left( B\right) =1$ and finally the joint system $\rho ^{AB}=\rho
^{A}\otimes \rho ^{B}=\frac{1}{2}\left| 00\right\rangle \left\langle
00\right| +\frac{1}{2}\left| 01\right\rangle \left\langle 01\right| $ has
entropy $S\left( A,B\right) =1.$

\item Introducing an auxiliary system $B$, whose state has an orthogonal
basis $\left| i\right\rangle $ corresponding to the states $\rho _{i}$ of
system $A.$ The joint system will have a state $\rho ^{AB}=\sum_{i}p_{i}\rho
_{i}\otimes \left| i\right\rangle \left\langle i\right| ,$ and its entropy
according to the joint entropy theorem (\ref{Joint_entropy_theorem}) is $%
S\left( A,B\right) =H\left( p_{i}\right) +\sum_{i}p_{i}\rho _{i}.$ Moreover
the entropy of each subsystem will be $S\left( A\right) =S\left(
\sum_{i}p_{i}\rho _{i}\right) $ and $S\left( B\right) =S\left(
\sum_{i}p_{i}\left| i\right\rangle \left\langle i\right| \right) =H\left(
p_{i}\right) ,$ and by applying subadditivity (property 3) $S\left(
A,B\right) \leq S\left( A\right) +S\left( B\right) ,$ q.e.d. Concerning its
equality conditions assume $\rho =\rho _{i}$ for every $i,$ then by
calculating $\rho ^{AB}=\sum_{i}p_{i}\rho _{i}\otimes \left| i\right\rangle
\left\langle i\right| =\rho \otimes \sum_{i}p_{i}\left| i\right\rangle
\left\langle i\right| =\rho ^{A}\otimes \rho ^{B},$ and equality follows
from (\ref{entropy_of_tensor_product}). Conversely if $S\left(
\sum_{i}p_{i}\rho _{i}\right) =\sum_{i}p_{i}S\left( \rho _{i}\right) ,$ and
suppose there is at least one density matrix which is not equal to the
others, say $\sigma \triangleq \rho _{j},$ then%
\begin{equation}
S\left( q\rho +p\sigma \right) =qS\left( \rho \right) +pS\left( \sigma
\right) ,  \label{S(q*rho+p*sigma)=q*S(rho)+p*S(sigma)}
\end{equation}%
where the following quantities where defined $q\triangleq \sum_{i\neq
j}p_{i},$ $p\triangleq p_{j}$ and $\rho \triangleq \rho _{i}$ for $i\neq j.$
If the density matrices $\rho ,\sigma $ have a spectral decomposition, say $%
\rho =\sum_{i}\lambda _{i}\left| i\right\rangle _{\rho }\left\langle
i\right| _{\rho },$ $\sigma =\sum_{j}\kappa _{j}\left| j\right\rangle
_{\sigma }\left\langle j\right| _{\sigma }.$ It is easy to find out that%
\begin{equation}
qS\left( \rho \right) +pS\left( \sigma \right) =\sum_{m}q\lambda _{m}\log
\left( q\lambda _{m}\right) +\sum_{m}p\kappa _{m}\log \left( p\kappa
_{m}\right) .  \label{right_side_of_S(q*rho+p*sigma)=q*S(rho)+p*S(sigma)}
\end{equation}%
This was the right hand side of (\ref{S(q*rho+p*sigma)=q*S(rho)+p*S(sigma)}%
). To get the left hand side assume that the matrix $q\rho +p\sigma $ has a
spectral decomposition $q\rho +p\sigma =\sum_{m}\mu _{m}\left|
m\right\rangle _{\rho \sigma },$ one can find unitary matrices connecting
these bases, $\left| i\right\rangle _{\rho }=\sum_{m}u_{im}\left|
m\right\rangle _{\rho \sigma }$ and $\left| j\right\rangle _{\sigma
}=\sum_{m}w_{im}\left| m\right\rangle _{\rho \sigma }$. This implies that%
\begin{eqnarray}
S\left( q\rho +p\sigma \right) &=&-\sum_{n}\left\langle n\right| _{\rho
\sigma }\left[ \left( q\sum_{iml}\lambda _{i}u_{im}\left| m\right\rangle
_{\rho \sigma }u_{il}^{\ast }\left\langle l\right| _{\rho \sigma
}+p\sum_{jml}\kappa _{j}w_{jm}\left| m\right\rangle _{\rho \sigma
}w_{jl}^{\ast }\left\langle l\right| _{\rho \sigma }\right) \right.
\label{left_side_of_S(q*rho+p*sigma)=q*S(rho)+p*S(sigma)} \\
&&\left. \times \log \left( q\sum_{iml}\lambda _{i}u_{im}\left|
m\right\rangle _{\rho \sigma }u_{il}^{\ast }\left\langle l\right| _{\rho
\sigma }+p\sum_{jml}\kappa _{j}w_{jm}\left| m\right\rangle _{\rho \sigma
}w_{jl}^{\ast }\left\langle l\right| _{\rho \sigma }\right) \right] \left|
n\right\rangle _{\rho \sigma }  \notag \\
&=&-\sum_{n}\left( q\sum_{iml}\lambda _{i}u_{im}u_{il}^{\ast
}+p\sum_{jml}\kappa _{j}w_{jm}w_{jl}^{\ast }\right) \log \left(
q\sum_{iml}\lambda _{i}u_{im}u_{il}^{\ast }+p\sum_{jml}\kappa
_{j}w_{jm}w_{jl}^{\ast }\right)  \notag \\
&=&-\sum_{n}\left( q\lambda _{n}+p\kappa _{n}\right) \log \left( q\lambda
_{n}+p\kappa _{n}\right) .  \notag
\end{eqnarray}%
In the last step the fact that $u_{ij}$ and $w_{ij}$ are unitary ($%
UU^{\dagger }=I\Rightarrow \sum_{j}u_{ij}u_{lj}^{\ast }=\delta _{il})$ was
employed. Taking the left and right hand side of equation (\ref%
{S(q*rho+p*sigma)=q*S(rho)+p*S(sigma)}) as found in (\ref%
{right_side_of_S(q*rho+p*sigma)=q*S(rho)+p*S(sigma)}) and (\ref%
{left_side_of_S(q*rho+p*sigma)=q*S(rho)+p*S(sigma)}) correspondingly, it is
simple to verify that%
\begin{equation*}
\sum_{m}q\lambda _{m}\log \left( 1+p\kappa _{m}\right) +\sum_{m}p\kappa
_{m}\log \left( 1+q\lambda _{m}\right) =0.
\end{equation*}%
The fact that $\log \left( 1+p\kappa _{m}\right) ,\log \left( 1+q\lambda
_{m}\right) \geq 0,$ implies that both summations are greater or equal to
zero, and the last equality leaves no other case than being both of them
zero. This can only happen if for the non-zero $\lambda _{m},$ $p_{m}$ are
null. An alternative proof that von Neumann entropy is concave, can \ be
presented by defining $f\left( p\right) \triangleq S\left( p\rho +\left(
1-p\right) \sigma \right) .$ Then by calculus if $f^{\prime \prime }\left(
p\right) \leq 0,$ then $f$ is concave, that is for $0\leq p\leq 1,$ $f\left(
px+\left( 1-p\right) y\right) \geq pf\left( x\right) +\left( 1-p\right)
f\left( y\right) .$ Selecting $x=1,$ $y=0,$ $f\left( p\right) \geq pf\left(
1\right) +\left( 1-p\right) f\left( 0\right) ,$ which according to the
definition of $f,$ implies that $S\left( p\rho +\left( 1-p\right) \sigma
\right) \geq pS\left( \rho \right) +\left( 1-p\right) S\left( \sigma \right)
.$

\item Refer to \cite[p.518,519]{NielsenChuang}.

\item For the proof refer to \cite[p.519-521]{NielsenChuang}. The fact that
these inequalities are equivalent, will be presented here. If $S\left(
R\right) +S\left( B\right) \leq S\left( R,C\right) +S\left( B,C\right) $
holds then by introducing an auxiliary system $A,$ which purifies the system 
$RBC,$ $S\left( R\right) =S\left( A,B,C\right) $ and $S\left( R,C\right)
=S\left( A,B\right) ,$ so the last inequality becomes $S\left( A,B,C\right)
+S\left( B\right) \leq S\left( A,B\right) +S\left( B,C\right) .$ Conversely
if $S\left( R,B,C\right) +S\left( B\right) \leq S\left( R,B\right) +S\left(
B,C\right) $ holds by inserting again a system $A$ purifying system $RBC,$
because $S\left( R,B,C\right) =S\left( A\right) $ and $S\left( R,B\right)
=S\left( A,C\right) ,$ the last inequality becomes $S\left( A\right)
+S\left( B\right) \leq S\left( A,C\right) +S\left( B,C\right) .$ From this
another equivalent form to write strong subadditivity is $0\leq S\left(
C|A\right) +S\left( C|B\right) $ or $S\left( A\right) +S\left( B\right)
-S\left( A,B\right) +S\left( A\right) +S\left( C\right) -S\left( A,C\right)
\leq 2S\left( A\right) \Leftrightarrow S\left( A:B\right) +S\left(
A:C\right) \leq 2S\left( A\right) .$ This inequality corresponds to the
second property of Shannon entropy, where $H\left( X:Y\right) \leq H\left(
X\right) ,$ which is not always true for quantum information theory. For
example, the composite system $\left| AB\right\rangle =\frac{1}{2}\left|
00\right\rangle +\frac{1}{2}\left| 11\right\rangle ,$ has entropy $S\left(
A,B\right) =0$ because it is pure, and $\rho ^{A}=\frac{1}{2}\left|
0\right\rangle \left\langle 0\right| +\frac{1}{2}\left| 1\right\rangle
\left\langle 1\right| ,$ thus $S\left( A\right) =1$ and similarly $S\left(
B\right) =1.$ Hence $S\left( A:B\right) >S\left( A\right) .$

\item Refer to \cite[p.523]{NielsenChuang}.

\item Refer to \cite[p.523]{NielsenChuang}.

\item Refer to \cite[p.523]{NielsenChuang}.

\item For the proof refer to \cite[p.520]{NielsenChuang}. Concerning this
property it should be emphasized that joint concavity implies concavity in
each input; this is obvious by selecting $B_{1}\triangleq B_{2}\triangleq B$
or $A_{1}\triangleq A_{2}\triangleq A.$ The converse is not true. For
example by choosing $f\left( x,y\right) =-x^{2}e^{y},$ which is convex on $x$
because $\frac{\partial }{\partial x^{2}}f=-2e^{y}\leq 0$ for every $x,y,$
and convex on $y$ because $\frac{\partial }{\partial y^{2}}f=-x^{2}e^{y}\leq
0$ for every $x,y.$ However $f\left( \frac{1}{3}4+\frac{2}{3}\frac{1}{3},%
\frac{1}{3}\left( -3\right) +\frac{2}{3}\frac{2}{3}\right) \simeq -0.57$
which is less than $\frac{1}{3}f\left( 4,\frac{1}{3}\right) +\frac{2}{3}%
f\left( -3,\frac{2}{3}\right) \simeq -0.41.$

\item Refer to \cite[p.524,525]{NielsenChuang}.
\end{enumerate}

\subsubsection{Using strict concavity to prove other properties of von
Neumann entropy}

Strict concavity (property 5) of von Neumann entropy, can be used to prove (%
\ref{S(rho)<=logd(=)}) and moreover that the completely mixed state $\frac{1%
}{d}I$ on a space of $d$ dimensions is the unique state of maximal entropy.
To do this the following result is stated: for a $d\times d$ normal matrix $%
A $ there exists a set of unitary matrices $U_{i}$ such that 
\begin{equation}
\sum_{i=1}^{d}U_{i}^{\left( A\right) }AU_{i}^{\left( A\right) \dagger
}=tr\left( A\right) I.  \label{equivalent_transformation_to_unit}
\end{equation}%
For a proof refer to appendix \ref{appendix_Special_diagonal_normal_matr}
(the proof of a more general proposition is due to Abbas Edalat). To prove
the uniqueness of $\frac{1}{d}I$ as a maximal state, take $A\equiv \rho $
any quantum state, then $S\left( \rho \right) =\frac{1}{d}%
\sum_{i=1}^{d}S\left( \rho \right) \overset{\text{property 2}}{=}%
\sum_{i=1}^{d}\frac{1}{d}S\left( U_{i}^{\left( \rho \right) }\rho
U_{i}^{\left( \rho \right) \dagger }\right) \overset{\text{property 5}}{\leq 
}S\left( \sum_{i=1}^{d}\frac{1}{d}U_{i}^{\left( \rho \right) }\rho
U_{i}^{\left( \rho \right) \dagger }\right) \overset{\left( \ref%
{equivalent_transformation_to_unit}\right) }{=}S\left( \frac{1}{d}I\right) .$
Hence by strict concavity (property 5) any state $\rho $ has less or equal
entropy to the completely mixed state of $\frac{1}{d}I,$ and in order to be
equal they should be identical.

Using von Neumann entropy a proof of the concavity of Shannon entropy can be
provided. Let $p_{i}$ and $q_{i}$ two probability distributions. Then 
\begin{equation}
H\left( \sum_{i}p_{i}q_{j}\right) =S\left( \sum_{i}p_{i}q_{j}\left|
j\right\rangle \left\langle j\right| \right) \overset{\text{property 5}}{%
\geq }\sum_{i}p_{i}S\left( qj\left| j\right\rangle \left\langle j\right|
\right) =\sum_{i}p_{i}H\left( q_{j}\right) ,
\label{concavity_of_Shn_entr_by_VonNeum}
\end{equation}%
with equality if and only if $q_{j}\left| j\right\rangle \left\langle
j\right| $s are the same, that is $q_{j}$s are identical.

\subsection{Measurement in quantum physics and information theory\label%
{subsec_Measurement_and_Info}}

As already noted in the introduction of the present section, quantum physics
seems very puzzling. This is because there are two types of evolution a
quantum system can undergo: unitary and measurement. One understands that
the first one is needed to preserve probability during evolution (see
subsection \ref{subsec_Basic_feat_of_quant_mech}). Then why a second type of
evolution is needed? Information theory can explain this, using the
following results:

\begin{enumerate}
\item \emph{Projective measurement can increase entropy.} This is derived
using strict concavity. Let $P$ be a projector and $Q\triangleq I-P$ the
complementary projector, then there exist unitary matrices $U_{1},$ $U_{2}$
and a probability $p$ such that for all $\rho $, $P\rho P+Q\rho Q=pU_{1}\rho
U_{1}^{\dagger }+\left( 1-p\right) U_{2}\rho U_{2}^{\dagger }$ (refer to \ref%
{appendix_Proj_measur_and_Unitary} for a proof), thus 
\begin{gather}
S\left( P\rho P+Q\rho Q\right) =S\left( pU_{1}\rho U_{1}^{\dagger }+\left(
1-p\right) U_{2}\rho U_{2}^{\dagger }\right) \overset{\text{property 5}}{%
\geq }pS\left( U_{1}\rho U_{1}^{\dagger }\right) +\left( 1-p\right) S\left(
U_{2}\rho U_{2}^{\dagger }\right)  \label{proj_measur_increas_entr} \\
\overset{\text{property 2}}{=}pS\left( \rho \right) +\left( 1-p\right)
S\left( \rho \right) =S\left( \rho \right) .  \notag
\end{gather}%
Because of strict concavity the equality holds if and only if $P\rho P+Q\rho
Q=\rho .$

\item \emph{General measurement can decrease entropy.} One can convince
himself by considering a qubit in state $\rho =\frac{1}{2}\left|
0\right\rangle \left\langle 0\right| +\frac{1}{2}\left| 1\right\rangle
\left\langle 1\right| ,$ which is not pure thus $S\left( \rho \right) >0,$
which is measured using the measurement matrices $M_{1}=\left|
0\right\rangle \left\langle 0\right| $ and $M_{2}=\left| 0\right\rangle
\left\langle 1\right| .$ If the result of the measurement is unknown then
the state of the system afterwards is $\rho ^{\prime }=M_{1}\rho
M_{1}^{\dagger }+M_{2}\rho M_{2}^{\dagger }=\left| 0\right\rangle
\left\langle 0\right| ,$ which is pure, hence $S\left( \rho ^{\prime
}\right) =0<S\left( \rho \right) .$

\item \emph{Unitary evolution preserves entropy}. This is already seen as
von Neumann's entropy property 2.
\end{enumerate}

Now one should remember the information theoretic interpretation of entropy
given throughout this chapter: entropy is the amount of knowledge one has
about a system. Result 3 instructs that if only unitary evolutions were
present in quantum theory, then no knowledge on any physical system could
exist! One is relieved by seeing that knowledge can decrease or increase by
measurements, as seen by results 1 and 2, and of course that is what
measurements were meant to be in the first place.

\chapter{Some important results of information theory}

Some very important results derived by information theory, which will be
useful for the development of the next two chapters, concern the amount of
accessible information and how data can be processed. These are presented
correspondingly in section \ref{sec_Accessible_Info} and in section \ref%
{sec_Data_Process}, both for the classical and the quantum case.

\section{Accessible information\label{sec_Accessible_Info}}

Information is not always perfectly known, for example during a transmission
over a noisy channel there is a possibility of information loss. This means
that obtaining upper bounds of accessible information can be very useful in
practice. These upper bounds are calculated in subsection \ref%
{subsec_Acces_class_info}\ for the classical and in subsection \ref%
{subsec_Access_quant_info} for the quantum case.

\subsection{Accessible classical information: Fano's inequality\label%
{subsec_Acces_class_info}}

Of major importance, in classical information theory, is the amount of
information that can be extracted from a random variable $X$ based on the
knowledge of another random variable $Y.$ That should be given as an upper
bound for $H\left( X|Y\right) ,$ and is going to be calculated next. Suppose 
$\tilde{X}\triangleq f\left( Y\right) $ is some function which is used as
the best guess for $X.$ Let $p_{e}\triangleq p\left( X\neq \tilde{X}\right) $
be the probability that this guess is incorrect. Then an 'error' random
variable can be defined%
\begin{equation*}
E\triangleq \left\{ 
\begin{array}{c}
1,\text{ }X\neq \tilde{X} \\ 
0,\text{ }X=\tilde{X}%
\end{array}%
\right. ,
\end{equation*}%
thus $H\left( E\right) =H\left( p_{e}\right) .$ Using the chaining rule for
conditional entropies $H\left( E,X|Y\right) =H\left( E|X,Y\right) +H\left(
X|Y\right) $ (Shannon entropy, property 8), however $E$ is completely
determined once $X$ and $Y$ are known, so $H\left( E|X,Y\right) =0$ and
hence 
\begin{equation}
H\left( E,X|Y\right) =H\left( X|Y\right) .  \label{H(E,X|Y)=H(X|Y)}
\end{equation}%
Applying chaining rule for conditional entropies (Shannon entropy, property
8) again, but for different variables 
\begin{equation}
H\left( E,X|Y\right) =H\left( X|E,Y\right) +H\left( E|Y\right) ,
\label{H(E,X|Y)=H(X|E,Y)+H(E|Y)}
\end{equation}%
and further because conditioning reduces entropy (Shannon entropy, property
7), $H\left( E|Y\right) \leq H\left( E\right) =H\left( p_{e}\right) ,$
whence by (\ref{H(E,X|Y)=H(X|Y)}) and (\ref{H(E,X|Y)=H(X|E,Y)+H(E|Y)}) 
\begin{equation}
H\left( X|Y\right) =H\left( E,X|Y\right) \leq H\left( X|E,Y\right) +H\left(
p_{e}\right) .  \label{fano_ineq_helping_relation}
\end{equation}%
Finally $H\left( X|E,Y\right) $ should be bounded as follows, after some
algebra%
\begin{eqnarray*}
H\left( X|E,Y\right) &=&p\left( E=0\right) H\left( X|E=0,Y\right) +p\left(
E=1\right) H\left( X|E=1,Y\right) \\
&\leq &p\left( E=0\right) \cdot 0+p_{e}\log \left( \left| X\right| -1\right)
=p_{e}\log \left( \left| X\right| -1\right) .
\end{eqnarray*}%
This relation with the help of (\ref{fano_ineq_helping_relation}) is known
as \emph{Fano's inequality}:%
\begin{equation}
H\left( p_{e}\right) +p_{e}\log \left( \left| X\right| -1\right) \geq
H\left( X|Y\right) .  \label{Fano's_inequality}
\end{equation}

\subsection{Accessible quantum information: quantum Fano inequality and the
Holevo bound\label{subsec_Access_quant_info}}

\subsubsection{Quantum Fano inequality}

There exists analogous relation to (\ref{Fano's_inequality}), in quantum
information theory, named \emph{quantum Fano inequality}:%
\begin{equation}
S\left( \rho ,\mathcal{E}\right) \leq H\left( F\left( \rho ,\mathcal{E}%
\right) \right) +\left( 1-F\left( \rho ,\mathcal{E}\right) \right) \log
\left( d^{2}-1\right) .  \label{quantum_Fano_inequality}
\end{equation}%
Where $F\left( \rho ,\mathcal{E}\right) $ is the entanglement fidelity of a
quantum operation defined in (\ref{fidelity_entangl}), for more details
refer to appendix \ref{appendix_Fidelity}. In the above equation, the \emph{%
entropy exchange}\ of the operation $\mathcal{E}$\ upon $\rho ,$ was
introduced%
\begin{equation*}
S\left( \rho ,\mathcal{E}\right) \triangleq S\left( R^{\prime },Q^{\prime
}\right) ,
\end{equation*}%
which is a measure of the noise caused by $\mathcal{E}$, on a quantum system 
$Q\ $($\rho \equiv \rho ^{Q}$), purified by $R$. The prime notation is used
to indicate the states after the application of $\mathcal{E}$. Note that the
entropy exchange, does not depend upon the way in which the initial state of 
$Q,$ is purified by $R.$ This is because any two purifications of $Q$ into $%
RQ$ are related by a unitary operation on the system $R,$ \cite[p.111]%
{NielsenChuang}, and because of von Neumann entropy, property 2.

Quantum Fano inequality (\ref{quantum_Fano_inequality}) is proven by taking
an orthonormal basis $\left| i\right\rangle $ for the system $RQ,$ chosen so
that the first state in the set $\left| 1\right\rangle =\left|
RQ\right\rangle .$ Forming the quantities $p_{i}\triangleq \left\langle
i\right| \rho ^{R^{\prime }Q^{\prime }}\left| i\right\rangle ,$ then it
follows from (\ref{proj_measur_increas_entr}) that $S\left( R^{\prime
},Q^{\prime }\right) \leq H\left( p_{1},\ldots ,p_{d^{2}}\right) ,$ and with
some simple algebra $H\left( p_{1},\ldots ,p_{d^{2}}\right) =H\left(
p_{1}\right) +\left( 1-p_{1}\right) H\left( \frac{p_{2}}{1-p_{1}},\ldots ,%
\frac{p_{d^{2}}}{1-p_{1}}\right) \leq \log \left( d^{2}-1\right) ,$ and
since $p_{1}\triangleq F\left( \rho ,\mathcal{E}\right) ,$ q.e.d.

\subsubsection{The Holevo bound}

Another result giving an upper bound of accessible quantum information is
the \emph{Holevo bound} \cite{Holevo}%
\begin{equation}
H\left( X:Y\right) \leq S\left( \rho \right) -\underset{x}{\sum }%
p_{x}S\left( \rho _{x}\right) ,  \label{Holevo_bound}
\end{equation}%
where $\rho =\sum_{x}p_{x}\rho _{x}.$ Moreover the right hand side of this
inequality is useful in quantum information theory, and hence it is given a
special name: \emph{Holevo }$\chi $\emph{\ quantity}. Concerning its proof
assume that someone, named $P,$ prepares some quantum information system $Q$
with states $\rho _{X}$, where $X=0,\ldots ,n,$ having probabilities $%
p_{0},\ldots ,p_{n}.$ The quantum information $Q$ is going to be measured by
another person, $M,$ using POVM elements $\left\{ E_{y}\right\} =\left\{
E_{0},\ldots ,E_{m}\right\} $ on the state and will have an outcome $Y.$ The
state of the total system before measurement will then be%
\begin{equation*}
\rho ^{PQM}=\underset{x}{\sum }p_{x}\left| x\right\rangle \left\langle
x\right| \otimes \rho _{x}\otimes \left| 0\right\rangle \left\langle
0\right| ,
\end{equation*}%
where the tensor product was in respect to the order $PQM.$ Matrix $\left|
0\right\rangle \left\langle 0\right| $ represents the initial state of the
measurement system, which holds before getting any information. The
measurement is described by an operator $\mathcal{E}$, that acts on each
state $\sigma $ of $Q$ by measuring it with POVM elements $\left\{
E_{y}\right\} ,$ and storing on $M$ the outcome. This can be expressed by%
\begin{equation*}
\mathcal{E}\left( \sigma \otimes \left| 0\right\rangle \left\langle 0\right|
\right) =\underset{y}{\sum }\sqrt{E_{y}}\sigma \sqrt{E_{y}}\otimes \left|
y\right\rangle \left\langle y\right| .
\end{equation*}%
Quantum operation $\mathcal{E}$ is trace preserving. To see this first
notice that it is made us of operations elements $\left\{ \sqrt{E_{y}}%
\otimes U_{y}\right\} ,$ where $U_{y}\left| y^{\prime }\right\rangle
\triangleq \left| y^{\prime }+y\right\rangle ,$ with $+$ the modulo $n+1$
addition. Of course $U_{y}$ is unitary since it is a map taking $\left|
y^{\prime }\right\rangle $ basis vector to another basis vector $\left|
y^{\prime }+y\right\rangle ,$ and hence it is change of basis from one basis
to a cyclic permutation of the same basis. Now because $E_{y}$ are POVM $%
I=\sum_{y}E_{y}=\sum_{y}\sqrt{E_{y}}^{\dagger }\sqrt{E_{y}}\overset{%
U_{y}^{\dagger }U_{y}=I}{\Rightarrow }\sum_{y}\sqrt{E_{y}}^{\dagger }\sqrt{%
E_{y}}\otimes U_{y}^{\dagger }U_{y}=I,$ q.e.d.

Subsequently primes are used to denote states after the application of $%
\mathcal{E}$, and unprimed notation for states before its application. Note
now that $S\left( P:Q\right) =S\left( P:Q,M\right) $ since $M$ is initially
uncorrelated with $P$ and $Q,$ and because by applying operator $\mathcal{E}$
it is not possible to increase mutual information (von Neumann entropy,
property 10) $S\left( P:Q,M\right) \geq S\left( P^{\prime }:Q^{\prime
},M^{\prime }\right) .$ Putting these results together%
\begin{equation}
S\left( P^{\prime }:M^{\prime }\right) \leq S\left( P:Q\right) .
\label{Holevo_inequality_to_by_proven}
\end{equation}%
The last equation, with a little algebra is understood to be the Holevo
bound. The one on the right of (\ref{Holevo_inequality_to_by_proven}) can be
found by thinking that $\rho ^{PQ}=\sum_{x}p_{x}\left| x\right\rangle
\left\langle x\right| \otimes \rho _{x},$ hence $S\left( P\right) =H\left(
p_{x}\right) ,$ $S\left( Q\right) =S\left( \rho \right) ,$ and $S\left(
P,Q\right) =H\left( p_{x}\right) +\sum_{x}p_{x}S\left( \rho _{\chi }\right) $
(von Neumann entropy, property 6), thus 
\begin{equation*}
S\left( P:Q\right) =S\left( P\right) +S\left( Q\right) -S\left( P,Q\right)
=S\left( \rho \right) -\sum_{x}p_{x}S\left( \rho _{x}\right) .
\end{equation*}%
Now the left hand side of (\ref{Holevo_inequality_to_by_proven}) is found by
noting that after a measurement 
\begin{equation*}
\rho ^{P^{\prime }Q^{\prime }M^{\prime }}=\sum_{xy}p_{x}\left|
x\right\rangle \left\langle x\right| \otimes \sqrt{E_{y}}\rho _{x}\sqrt{E_{y}%
}\otimes \left| y\right\rangle \left\langle y\right| ,
\end{equation*}%
tracing out the system $Q^{\prime }$ and using the observation that the
joint distribution $p\left( x,y\right) $ for the pair $\left( X,Y\right) $
satisfies $p\left( x,y\right) =p_{x}p\left( y|x\right) =p_{x}$tr$\left( \rho
_{x}E_{y}\right) =p_{x}$tr$\left( \sqrt{E_{y}}\rho _{x}\sqrt{E_{y}}\right) ,$
it is straightforward to see that $\rho ^{P^{\prime }M^{\prime
}}=\sum_{xy}p\left( x,y\right) \left| x\right\rangle \left\langle x\right|
\otimes \left| y\right\rangle \left\langle y\right| ,$ whence 
\begin{equation*}
S\left( P^{\prime }:M^{\prime }\right) =S\left( P^{\prime }\right) +S\left(
M^{\prime }\right) -S\left( P^{\prime },M^{\prime }\right) =H\left( X\right)
+H\left( Y\right) -H\left( X,Y\right) =H\left( X:Y\right) ,
\end{equation*}%
q.e.d.

\section{Data processing\label{sec_Data_Process}}

As it is widely known, information except of being stored and transmitted,
it is also processed. In subsection \ref{subsec_Class_data_proc} data
processing is defined for the classical case and the homonymous inequality
is proven. An analogous definition and inequality are demonstrated for the
quantum case in subsection \ref{subsec_Quant_data_process}.

\subsection{Classical data processing\label{subsec_Class_data_proc}}

Classical data processing can be described in mathematical terms by a chain
of random variables 
\begin{equation}
X_{1}\rightarrow X_{2}\rightarrow X_{3}\rightarrow \cdots \rightarrow X_{n}
\label{classical_data_processing}
\end{equation}

where $X_{i}$ is the $i$-th step of processing. Of course each step depends
only from the information gained by the previous, that is $p\left(
X_{n+1}=x_{n+1}|X_{n}=x_{n},\ldots ,X_{1}\right) =p\left(
X_{n+1}=x_{n+1}|X_{n}=x_{n}\right) ,$ which defines a Markov chain. But as
it is already accentuated information is a physical entity which can be
distorted by noise. Thus if $X\rightarrow Y\rightarrow Z$ is a Markov chain
representing an information process one can prove 
\begin{equation}
H\left( X\right) \geq H\left( X:Y\right) \geq H\left( X:Z\right) ,
\label{def_data_proc_ineq}
\end{equation}%
which is known as the \emph{data processing inequality}. This inequality
reveals a mathematical insight of the following physical truth: if a system
described by a random variable $X$ is subjected to noise, producing $Y,$
then further data process cannot be used to increase the amount of mutual
information between the output process and the original information $X;$
once information is lost, it cannot be restored. It is worth mentioning that
in a data process chain $X\rightarrow Y\rightarrow Z,$ information a system $%
Z$ shares with $X$ must be information which $Z$ also shares with $Y;$ the
information is 'pipelined' from $X$ through $Y$ to $Z.$ This is described by
the \emph{data pipelining inequality}%
\begin{equation*}
H\left( Z:Y\right) \geq H\left( Z:X\right) .
\end{equation*}%
This is derived by (\ref{def_data_proc_ineq}) and noting that $X\rightarrow
Y\rightarrow Z$ is a Markov chain, if and only if 
\begin{gather*}
p\left( Z=z|Y=y,X=x\right) =p\left( Z=z|Y=y\right) \Leftrightarrow \frac{%
p\left( X=x,Y=y,Z=z\right) }{p\left( X=x,Y=y\right) }=\frac{p\left(
Y=y,Z=z\right) }{p\left( Y=y\right) }\Leftrightarrow \\
\frac{p\left( X=x,Y=y,Z=z\right) }{p\left( Y=y,Z=z\right) }=\frac{p\left(
X=x,Y=y\right) }{p\left( Y=y\right) }\Leftrightarrow p\left(
X=x|Y=y,Z=z\right) =p\left( X=x|Y=y\right) ,
\end{gather*}%
if and only if, $Z\rightarrow Y\rightarrow X$ is a Markov chain too. In the
above proof there is no problem with null probabilities in the denominators,
because then every probability would be null, and the proven result would
still hold.

\subsection{Quantum data processing\label{subsec_Quant_data_process}}

The quantum analogue of data processing (\ref{classical_data_processing}) is
described by a chain of quantum operations%
\begin{equation*}
\rho \rightarrow \mathcal{E}_{1}\left( \rho \right) \rightarrow \left( 
\mathcal{E}_{2}\circ \mathcal{E}_{1}\right) \left( \rho \right) \rightarrow
\cdots \rightarrow \left( \mathcal{E}_{n}\circ \cdots \circ \mathcal{E}%
_{2}\circ \mathcal{E}_{1}\right) \left( \rho \right) .
\end{equation*}%
In the above model each step of process is obtained by application of a
quantum operator. By defining the \emph{coherent information}, 
\begin{equation}
I\left( \rho ,\mathcal{E}\right) \triangleq S\left( \mathcal{E}\left( \rho
\right) \right) -S\left( \rho ,\mathcal{E}\right) .
\label{def_coherent_info}
\end{equation}%
It can be proven that%
\begin{equation}
S\left( \rho \right) \geq I\left( \rho ,\mathcal{E}_{1}\right) \geq I\left(
\rho ,\mathcal{E}_{1}\circ \mathcal{E}_{2}\right) ,
\label{quantum_data_process}
\end{equation}%
which corresponds to the classical data processing inequality (\ref%
{def_data_proc_ineq}).

\chapter{Real-life applications of information theory}

After a theoretical presentation of information theory one should look over
its real-life applications. Two extremely useful applications of information
theory are data compression discussed in section \ref{sec_Data_Compress} and
transmission of information over noisy channels is the main topic of section %
\ref{sec_Noisy_Coding}.

\section{Data compression\label{sec_Data_Compress}}

Nowadays data compression is a widely applied procedure; everybody uses .zip
archives, listen to .mp3 music, watches videos in .mpeg format and exchanges
photos in .jpg files. Although in all these cases, special techniques
depending on the type of data are used, the general philosophy underlying
data compression is inherited by Shannon's noiseless channel coding theorem %
\cite{Shannon,ShannonWeaver} discussed in subsection \ref%
{subsec_Shan_Data_Compress}. In the quantum case data compression was
theoretically found to be possible in 1995 by Schumacher \cite{Schumacher}
and is presented in subsection \ref{subsec_Schumach_Quant_Data_Compress}.

\subsection{Shannon's noiseless channel coding theorem\label%
{subsec_Shan_Data_Compress}}

Shannon's main idea was to estimate the physical resources needed to
represent an information content, which as already seen in section \ref%
{sec_Classical_physics_&_info_th}\ are related to entropy. As a simple
example to understand how the theorem works, consider a binary alphabet to
be compressed. In this alphabet assume that 0 occurs with probability $1-p$
and 1 with probability $p.$ Then if $n$-bits strings are formed, they will
contain $\left( 1-p\right) n$ zero bits and $pn$ one bits, with very high
probability related to the magnitude of $n,$ according to the law of large
numbers. Since from all the possible strings to be formed these are the most
likely they are usually named \emph{typical sequences}. Calculating all the
combinations of ones and zeros, there are totally $\binom{n}{np}$ such
strings. Using Stirling's approximation $n!\simeq \left( \frac{n}{e}\right)
^{n},$ one finds that%
\begin{eqnarray}
\binom{n}{np} &=&\frac{n!}{\left( np\right) !\left[ n\left( 1-p\right) %
\right] !}\simeq \frac{\left( \frac{n}{e}\right) ^{n}}{\left( \frac{np}{e}%
\right) ^{np}\left( \frac{n\left( 1-p\right) }{e}\right) ^{n\left(
1-p\right) }}  \label{Shannon_data_compress_intuitive_proof_for_2} \\
&=&\frac{n^{n}}{\left( np\right) ^{np}\left( n\left( 1-p\right) \right)
^{n\left( 1-p\right) }}=2^{n\log n-np\log np-n\left( 1-p\right) \log n\left(
1-p\right) }  \notag \\
&=&2^{n\left[ -p\log p-\left( 1-p\right) \log \left( 1-p\right) \right]
}=2^{nH\left( p\right) }.  \notag
\end{eqnarray}%
Generalizing the above argument for an alphabet of $k$ letters $x_{i}\in X$
with occurrence probabilities $p\left( x_{i}\right) ,$ the number of typical
sequences can be easily calculated using combinatorics as before, and found
to be%
\begin{equation}
\binom{n}{np\left( x_{1}\right) ,np\left( x_{2}\right) ,\ldots ,np\left(
x_{k-1}\right) }=\frac{n!}{\underset{x\in X\backslash \left\{ x_{k}\right\} }%
{\prod }\left( np\left( x\right) \right) !}\simeq 2^{nH\left( X\right) }.
\label{Shannon_data_compress_intuitive_proof_for_n}
\end{equation}%
Where letters need not just be one symbol but also a sequence of symbols
like words of English language. Obviously the probability of such a sequence
to occur is approximately $2^{-nH\left( X\right) }.$ This approximate
probability gives a very intuitive way of understanding the mathematical
terminology of Shannon's noiseless theorem, where a sequence $%
y=x_{i_{1}}x_{i_{2}}x_{i_{3}}\cdots x_{i_{n}}$ is defined to be $\epsilon $%
-typical if the probability of its occurrence is%
\begin{equation*}
2^{-n\left( H\left( X\right) +\epsilon \right) }\leq p\left( y\right) \leq
2^{-n\left( H\left( X\right) -\epsilon \right) }.
\end{equation*}%
The set of all such sequences is denoted $T\left( n,\epsilon \right) .$
Another very useful way of writing this result is%
\begin{equation}
\left| \frac{1}{n}\log \left( \frac{1}{p\left(
x_{i_{1}}x_{i_{2}}x_{i_{3}}\cdots x_{i_{n}}\right) }\right) -H\left(
X\right) \right| \leq \epsilon .  \label{2nd_def_e-typical_sequences}
\end{equation}%
Now the theorem of typical sequences can be stated

\begin{quotation}
\textbf{Theorem of typical sequences:}
\end{quotation}

\begin{enumerate}
\item Fix $\epsilon >0,$ then for any $\delta >0,$ for sufficiently large $%
n, $ the probability that a sequence is $\epsilon $-typical is at least $%
1-\delta .$

\item For any fixed $\epsilon >o$ and $\delta >0,$ for sufficiently large $%
n, $ $\left( 1-\delta \right) 2^{n\left( H\left( X\right) -\epsilon \right)
}\leq \left| T\left( n,\epsilon \right) \right| \leq 2^{n\left( H\left(
X\right) +\epsilon \right) }.$

\item Let $S\left( n\right) $ be the collection of size at most $2^{nR},$ of
length $n$ sequences, where $R<H\left( X\right) $ is fixed. Then for any $%
\delta >0$ and for sufficiently large $n,$ $\sum_{y\in S\left( n\right)
}p\left( y\right) \leq \delta .$
\end{enumerate}

The above theorem is proven by using the law of large numbers. Moreover
Shannon's noiseless channel coding theorem, is just an application of the
last stated theorem. Shannon implemented a \emph{compression scheme} which
is just a map of an $n$-bit sequence $y=x_{i_{1}}x_{i_{2}}x_{i_{3}}\cdots
x_{i_{n}}$ to another one of $nR$-bits denoted by $C_{n}\left( y\right) .$
Of course in such a compression scheme an invert map $D_{n}$ ($D_{n}\circ
C_{n}=id_{X^{n}}$) should exist, which naturally would be named \emph{%
decompression scheme}. However the set of typical sequences, in non-trivial
cases, is only a subset of all the possible sequences and this drives to
failure of the schemes, when they will be invited to map the complementary
subset, known as \emph{atypical sequences}\ subset. This way further
nomenclature may be added by saying that a compression decompression scheme $%
\left( C_{n},D_{n}\right) $ is said to be \emph{reliable} if the probability
that $D_{n}\left( C_{n}\left( y\right) \right) =y$ approaches one as $n$
approaches infinity$.$ It is time to state the theorem.

\begin{quotation}
\textbf{Shannon's noiseless channel coding theorem:}

Assume an alphabet $X$ then if $R>H\left( X\right) $ is chosen there exists
a reliable compression decompression scheme of rate $R.$ Conversely, if $%
R<H\left( X\right) $ any such scheme will not be reliable.
\end{quotation}

This theorem is revealing a remarkable operational interpretation for the
entropy rate $H\left( X\right) $: it is just the minimal physical resources
necessary and sufficient to reliably transmit data. Finally it should be
stressed that somebody can have a perfectly reliable compression
decompression scheme just by extending maps to atypical sequences; in this
case there is just high probability that no more than $nR$ resources are
needed to carry the information.

\subsection{Schumacher's quantum noiseless channel coding theorem\label%
{subsec_Schumach_Quant_Data_Compress}}

It is yet quite surprising that quantum information can be compressed as was
proven by Schumacher \ref{subsec_Schumach_Quant_Data_Compress}. Assume that
quantum information transmitted can be in states $\left| x_{i}\right\rangle
\in H^{\otimes n}$ with probability $p\left( x_{i}\right) .$ This is
described by the density matrix $\rho =\sum_{i=1}^{n}p\left( x_{i}\right)
\left| x_{i}\right\rangle \left\langle x_{i}\right| .$ A
compression-decompression scheme of rate $R$ consists of two quantum
operations $\mathcal{C}_{n}$ and $\mathcal{D}_{n}$ analogous to the maps
defined for the classical case. The compression operation $\mathcal{C}_{n}$
is taking states from $H^{\otimes n}$ to $H^{\otimes nR}$ and the
decompression $\mathcal{D}_{n}$ returns them back, as Figure \ref%
{fig_Schumacher_coding} demonstrates. One can define a sequence $%
x_{i_{1}}x_{i_{2}}x_{i_{3}}\cdots x_{i_{n}}$ as $\epsilon $-typical by a
relation resembling to the classical (\ref{2nd_def_e-typical_sequences})%
\begin{equation*}
\left| \frac{1}{n}\log \left( \frac{1}{p\left( x_{i_{1}}\right) p\left(
x_{i_{2}}\right) \cdots p\left( x_{i_{n}}\right) }\right) -S\left( \rho
\right) \right| \leq \epsilon .
\end{equation*}%
A state $\left| x_{i_{1}}\right\rangle \left| x_{i_{2}}\right\rangle \cdots
\left| x_{i_{n}}\right\rangle $ is said to be $\epsilon $-typical if the
sequence $x_{i_{1}}x_{i_{2}}x_{i_{3}}\cdots x_{i_{n}}$ is $\epsilon $%
-typical. The $\epsilon $-typical subspace will be noted $T\left( n,\epsilon
\right) $ and the projector onto this subspace will be%
\begin{equation*}
P\left( n,\epsilon \right) =\underset{x_{i_{1}}x_{i_{2}}x_{i_{3}}\cdots
x_{i_{n}}\in T\left( n,\epsilon \right) }{\sum }\left|
x_{i_{1}}\right\rangle \left\langle x_{i_{1}}\right| \otimes \left|
x_{i_{2}}\right\rangle \left\langle x_{i_{2}}\right| \otimes \cdots \otimes
\left| x_{i_{n}}\right\rangle \left\langle x_{i_{n}}\right| .
\end{equation*}%
Now the quantum typical sequences theorem can be stated.

\begin{quotation}
\textbf{Typical subspace theorem:}
\end{quotation}

\begin{enumerate}
\item Fix $\epsilon >0$ then for any $\delta >0,$ for sufficiently large $n,$
tr$\left( P\left( n,\epsilon \right) \rho ^{\otimes n}\right) \geq 1-\delta
. $

\item For any fixed $\epsilon >o$ and $\delta >0,$ for sufficiently large $%
n, $ $\left( 1-\delta \right) 2^{n\left( S\left( X\right) -\epsilon \right)
}\leq \left| T\left( n,\epsilon \right) \right| \leq 2^{n\left( S\left(
X\right) +\epsilon \right) }.$

\item Let $S\left( n\right) $ be a projector onto any subspace of $%
H^{\otimes n}$ of dimension at most $2^{nR},$ where $R<S\left( \rho \right) $
is fixed. Then for any $\delta >0$ and for sufficiently large $n,$ tr$\left(
S\left( n\right) \rho ^{\otimes n}\right) \leq \delta .$
\end{enumerate}

Following the same principles the quantum version of Shannon's theorem as
proved by Schumacher is,

\begin{quotation}
\textbf{Schumacher's noiseless channel coding theorem:}

Let $\rho $ be information belonging in some a Hilbert space $H$ then if $%
R>S\left( \rho \right) $ there exists a reliable compression scheme.
Conversely if $R<S\left( \rho \right) $ any compression scheme is not
reliable.
\end{quotation}

The\ compression\ scheme found by Schumacher is%
\begin{equation*}
\mathcal{C}_{n}\left( \sigma \right) \triangleq P\left( n,\epsilon \right)
\sigma P\left( n,\epsilon \right) +\left| 0\right\rangle \left\langle
i\right| \sigma \left| i\right\rangle \left\langle 0\right| ,
\end{equation*}%
where $\left| i\right\rangle $ is an orthonormal basis for the
orthocomplement of the typical subspace, and $\left| 0\right\rangle $ is
some standard state. As one can see this quantum operation takes any state $%
\sigma $ from $H^{\otimes n}$ to $H^{\otimes nR},$ the subspace of $\epsilon 
$-typical sequences if $\sigma $ can be compressed, and if not it gives as
outcome the standard state $\left| 0\right\rangle ,$ which is meant to be a
failure. Finally $\mathcal{D}_{n}$ was found to be the identity map on $%
H^{\otimes nR},$ which obviously maps any compressed state back to $%
H^{\otimes nR}\leq H^{\otimes n}.$

\section{Information over noisy channels\label{sec_Noisy_Coding}}

It is an everyday life fact that communication channels are imperfect and
are always subject to noise which distorts transmitted information. This of
course prevents reliable communication without some special control of
information transmitted and received. One can use error correction in order
to achieve such a control, which is summarized in subsection \ref%
{subsec_Error_correction}. However there are some general theoretical
results concerning such transmissions, which help calculate the capacity of
noisy channels. The cases of classical information over noisy classical and
quantum channels each presented in subsections \ref{subsec_Noisy_Coding} and %
\ref{subsec_Class_over_noisy_Quant_ch}. A a summary for the up today results
for quantum information over noisy quantum channels is given in subsection %
\ref{subsec_Quant_info_over_Qch}.

\subsection{Error correction\label{subsec_Error_correction}}

Error correction is a practical procedure for transmission of information
over noisy channels. The intuitive idea behind it is common in every day
life. As an example recall a typical telephone conversation. If the
connection is of low quality, people communicating often need to repeat
their words, in order to protect their talk against the noise. Moreover
sometimes during a telephonic communication, one is asked to spell a word.
Then by saying words whose initials are the letters to be spelled,
misunderstanding is minimized. If someone wants to spell the word \emph{%
''phone''}, he can say \emph{''Parents''}, \emph{''Hotel''}, \emph{''Oracle''%
}, \emph{''None''} and \emph{''Evangelist''}. If instead of saying \emph{%
''None''}, he said \emph{''New''} the person at the other side of the line
could possibly hear \emph{''Mew''}. This example demonstrates why words
should be carefully selected. One can see that the last two words differ by
only one letter, their Hamming distance is small (refer to appendix \ref%
{appendix_Hamming} for a definition), hence they should select words with
higher distances.

The error correction procedure, as presented in the last paragraph, is the
encoding and transmission of a message longer than the one willing to
communicate, containing enough information to reconstruct the initial
message up to a probability. If one wish to encode a $k$-bit word $x$, into
an $n$-bit $y$ ($k\leq n$), then this encoding-decoding scheme is named $%
\left[ n,k\right] ,$ and represented by function $C\left( x\right) =y.$ In
the last case it is often written $x\in C.$ One can avoid misunderstandings
similar to \emph{''New''} and \emph{''Mew''}, as found in the above
paragraph, by asking each codeword to be of Hamming distance greater than $%
d. $ Then after receiving a codeword $y$, he tries to find to which Hamming
sphere, Sph$\left( x,d\right) ,$ it belongs, and then identifies the
received codeword with the center of the sphere: $x.$ Such a code is denoted
by $\left[ n,k,d\right] .$

The basic notions of classical and quantum error correction are summarized
in next paragraphs.

\subsubsection{Classical error correction}

From all the possible error correction codes, a subset is going to be
presented here, the linear ones. A member of this subset, namely a $\left[
n,k\right] $ linear code, is modeled by an $n\times l$ matrix $G,$ often
called the generator. The $k$-bit message $x,$ is treated as a column
vector, and the encoded $n$-bit message is the $Gx,$ where the numbers in
both $G$ and $x$ are numbers of $\mathbb{Z}_{2},$ that is zeros and ones,
and all the operations are performed modulo $2.$

The linear codes are used because for the case of $\left[ n,k\right] ,$ $nk$
bits are needed to represent it. In a general code $C,$ an $n$-bit string
would correspond to one of $2^{k}$ words, and to do this a table of $n2^{k}$
bits is needed. In contrast by using linear codes much memory is saved the
encoding program is more efficient.

In order to perform error correction one takes an $\left( n-k\right) \times
n $ matrix $H,$ named \emph{parity check}, having the property $HG=0.$ Then
for every codeword $y=Gx$ it is obvious that $Hy=0.$ Now if an noise was
present during transmission, one receives a state $y^{\prime }\triangleq
y+e, $ where $e$ is the error occurred. Hence $Hy^{\prime }\triangleq
Hy+He=He.$ Usually $Hy^{\prime }$ is called the \emph{error syndrome}. From
the error syndrome one can identify the initial $y$ if $t\geq d\left(
y^{\prime },y\right) =d\left( y+e,y\right) =d\left( e,0\right) ,$ and then
checking in which sphere $y^{\prime }\in $ Sph$\left( y,d\right) .$ To do
this the distance of the code must be defined by $d\equiv d\left( C\right)
\triangleq \underset{x,y\in C,x\neq y}{\min }d\left( x,y\right) ,$ that is
the spheres of radius $d$ must be distinct. Then if $d\geq 2t+1,$ up to $t$
bits can be corrected. All this are under the assumption that the
probability that the channel flips a bit is less than $\frac{1}{2}.$

It is easy to check that linear codes $\left[ n,k,d\right] ,$ must satisfy
the \emph{Singleton bound}%
\begin{equation}
n-k\geq d-1.  \label{singleton_bound}
\end{equation}%
One can further prove, that for large $n$ there exists an $\left[ n,k\right] 
$ error-correcting code, protecting against $t$ bits for some $k,$ such that 
\begin{equation}
\frac{k}{n}\geq 1-H_{\text{bin}}\left( \frac{t}{n}\right) .
\label{Gilbert-Varshamov_bound}
\end{equation}%
This is known as \emph{Gilbert-Varshamov bound}.

Some further definitions are needed. Suppose an $\left[ n,k\right] $ code $C$
is given, then its $\emph{dual}$ is denoted $C^{\perp },$ and has as
generator matrix $H^{T}$ and parity check $G^{T}.$ Thus the words in $%
C^{\perp }$ are orthogonal to $C.$ A code is said to be \emph{weakly
self-dual} if $C\subseteq C^{\perp },$ and \emph{strictly self dual} if $%
C=C^{\perp }.$

\subsubsection{Quantum error correction}

In what concerns quantum information theory, errors occurring are not of the
same nature as in the classical case. One has to deal, except from bit flip
errors, with phase flip errors. The first codes found to be able to both of
them are named \emph{Calderbank-Shor-Steane} after their inventors \cite%
{CalderbankShor,Steane96}. Assume $C_{1}$ and $C_{2}$ are $\left[ n,k_{1}%
\right] $ and $\left[ n,k_{2}\right] $ classical linear codes such that $%
C_{1}\subset C_{2}$ and both $C_{1}$ and $C_{2}^{\perp }$ can correct $t$
errors. Then an $\left[ n,k_{1}-k_{2}\right] $ quantum code CSS$\left(
C_{1},C_{2}\right) $ is defined, capable of correcting errors on $t$ qubits,
named the \emph{CSS code of }$C_{1}$\emph{\ over }$C_{2},$ via the following
construction. Suppose $x\in C_{1},$ then define $\left| x+C_{2}\right\rangle
\triangleq \frac{1}{\sqrt{\left| C_{2}\right| }}\sum_{y\in C_{2}}\left|
x+y\right\rangle ,$ where the addition is modulo $2.$ If now $x^{\prime }$
is an element of $C_{1}$ \ such that $x-x^{\prime }\in C_{2},$ then it easy
to verify that $\left| x+C_{2}\right\rangle =\left| x^{\prime
}+C_{2}\right\rangle ,$ and hence $\left| x+C_{2}\right\rangle $ depends
only upon the coset $C_{1}/C_{2}\ni x.$ Furthermore, if $x$ and $x^{\prime }$
belong to different coset of $C_{2},$ then for no $y,y^{\prime }\in C_{2}$
does $x+y=x^{\prime }+y^{\prime },$ and therefore $\left\langle
x+C_{2}|x^{\prime }+C_{2}\right\rangle =0.$ The quantum code CSS$\left(
C_{1},C_{2}\right) $ is defined to be the vector space spanned by the states 
$\left| x+C_{2}\right\rangle $ for all $x\in C_{1}.$ The number of cosets of 
$C_{2}$ in $C_{1}$ is $\frac{\left| C_{1}\right| }{\left| C_{2}\right| },$
so $\dim \left( \text{CSS}\left( C_{1},C_{2}\right) \right) =\frac{\left|
C_{1}\right| }{\left| C_{2}\right| }=2^{k_{1}-k_{2}},$ and therefore CSS$%
\left( C_{1},C_{2}\right) $ is an $\left[ n,k_{1}-k_{2}\right] $ quantum
code.

The quantum error correction can be exploited by the classical error
correcting properties of $C_{1}$ and $C_{2}^{\perp }.$ In the quantum case
there is like in the classical a possibility of a flip bit error, given in $%
e_{1},$ but additionally a phase flip error in denoted here by $e_{2}.$ Then
because of noise the original state $\left| x+C_{2}\right\rangle \left|
0\right\rangle $ has changed to $\frac{1}{\sqrt{\left| C_{2}\right| }}%
\sum_{y\in C_{2}}\left( -1\right) ^{\left( x+y\right) \cdot e_{2}}\left|
x+y+e_{1}\right\rangle \left| 0\right\rangle ,$ where an ancilla system $%
\left| 0\right\rangle $ to store the syndrome for the $C_{1}$ is introduced.
Applying the parity matrix to the deformed state $\frac{1}{\sqrt{\left|
C_{2}\right| }}\sum_{y\in C_{2}}\left( -1\right) ^{\left( x+y\right) \cdot
e_{2}}\left| x+y+e_{1}\right\rangle \left| H_{1}e_{1}\right\rangle ,$ and by
measuring the ancilla system the error syndrome is obtained and one corrects
the error by applying \texttt{NOT} gates to the flipped bits as indicated by 
$e_{1},$ giving $\frac{1}{\sqrt{\left| C_{2}\right| }}\sum_{y\in
C_{2}}\left( -1\right) ^{\left( x+y\right) \cdot e_{2}}\left|
x+y\right\rangle .$ If to this state Hadamard gates to each qubit are
applied, phase flip error are detected. One then takes $\frac{1}{\sqrt{%
\left| C_{2}\right| 2^{n}}}\sum_{z}\sum_{y\in C_{2}}\left( -1\right)
^{\left( x+y\right) \cdot \left( e_{2}+z\right) }\left| z\right\rangle ,$
and defining $z^{\prime }\triangleq z+e_{2},$ one can write the last state
as $\frac{1}{\sqrt{\left| C_{2}\right| 2^{n}}}\sum_{z^{\prime }}\sum_{y\in
C_{2}}\left( -1\right) ^{\left( x+y\right) \cdot z^{\prime }}\left|
z^{\prime }+e_{2}\right\rangle .$ Assuming $z^{\prime }\in C_{2}^{\perp }$
then it easy to verify that $\sum_{y\in C_{2}}\left( -1\right) ^{y\cdot
z^{\prime }}=\left| C_{2}\right| ,$ while if $z^{\prime }\notin C_{2}^{\perp
}$ then $\sum_{y\in C_{2}}\left( -1\right) ^{y\cdot z^{\prime }}=0.$ Hence
the state is further rewritten as $\sqrt{\frac{\left| C_{2}\right| }{2^{n}}}%
\sum_{z^{\prime }\in C_{2}^{\perp }}\left( -1\right) ^{x\cdot z^{\prime
}}\left| z^{\prime }+e_{2}\right\rangle ,$ which resembles to a bit flip
error described by vector $e_{2}.$ Following the same procedure for the bit
flips the state is finally as desired quantum error-corrected.

A quantum analogue of Gilbert-Varshamov bound is proven for the CSS codes,
guaranteeing the existence of good quantum codes. In the limit $n$ becomes
large, an $\left[ n,k\right] $ quantum code protecting up to $t$ errors
exist for some $k$ such that $\frac{k}{n}\geq 1-2H_{\text{bin}}\left( \frac{%
2t}{n}\right) .$

Concluding this summary on error correction, it is useful for quantum
cryptography to define codes by 
\begin{equation}
\left| x+C_{2}\right\rangle \triangleq \frac{1}{\sqrt{\left| C_{2}\right| }}%
\sum_{y\in C_{2}}\left( -1\right) ^{u\cdot v}\left| x+y\right\rangle ,
\label{CSS_quant_error_corr_used_in_QKD}
\end{equation}%
parametrized by $u$ and $v,$ and named CSS$_{u,v}\left( C_{1},C_{2}\right) ,$
which are equivalent to CSS$\left( C_{1},C_{2}\right) .$

\subsection{Classical information over noisy classical channels\label%
{subsec_Noisy_Coding}}

The theoretical study of transmission of information over noisy channels is
motivated by Shannon's corresponding theorem, which demonstrates the
existence of codes capable of realizing it, without giving clues how they
could be constructed. To model transmission of information over noisy
channel a finite \emph{input alphabet} $X$ and a finite \emph{output alphabet%
} $Y$ are considered; if a letter $x\in X$ is transmitted by one side, over
the noisy channel, then a letter $y\in Y$ is received by the other, with
probability $p\left( y|x\right) ,$ where of course $\sum_{y}p\left(
y|x\right) =1$ for all $x.$ The channel will be assumed \emph{memoryless} in
the sense of Markov's process, where the action on the currently transmitted
letter is independent of the previous one.

Now the process of transmitting information according to Shannon's noisy
channel coding theorem uses the result of the noiseless one. According to
this theorem it is always possible to pick up a reliable compression
decompression scheme of rate $R.$ Then a message $M$ can be viewed as one of
the possible $2^{nR}$ typical strings and encoded using the map $%
C_{n}:\left\{ 1,\ldots ,2^{nR}\right\} \rightarrow X^{n}$ which assigns $M$
to each $n$-sequence of the input alphabet. This sequence is sent over the
noisy channel and decoded using the map $D_{n}:Y^{n}\rightarrow \left\{
1,\ldots ,2^{nR}\right\} .$ This procedure is shown in Figure \ref%
{fig_Noisy_channel_coding}. It is very natural for a given encoding-decoding
pair to define the \emph{probability of error} as the maximum probability
over all messages $M$ that the decoded output of the channel is not equal to 
$M,$%
\begin{equation*}
p\left( C_{n},D_{n}\right) \triangleq \underset{M}{\max }\text{ }p\left(
D_{n}\left( Y\right) \neq M|X=C_{n}\left( M\right) \right) .
\end{equation*}%
Then it is said that a rate $R$ is \emph{achievable} if there exists such a
sequence of encoding-decoding pairs $\left( C_{n},D_{n}\right) $ and require
in addition $p\left( C_{n},D_{n}\right) $ approaching zero as $n$ approaches
infinite. The \emph{capacity} $C\left( \mathcal{N}\right) $ of a noisy
channel $\mathcal{N}$ is defined as the supremum over all the achievable
rates for the channel and is going to be calculated in the following
paragraph.

\begin{figure}[tbh]
\centering\includegraphics[width=15cm]{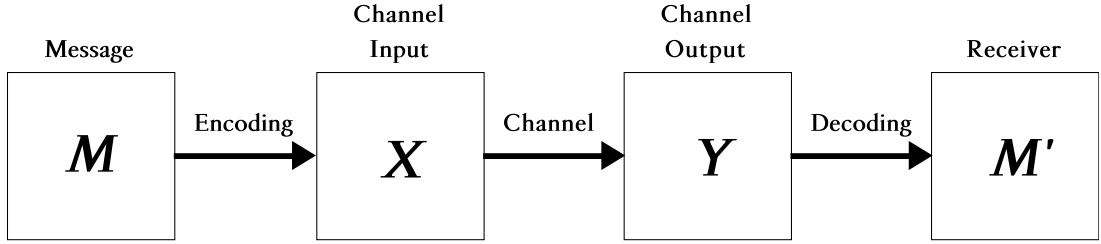}
\caption{The noisy channel coding problem for classical messages. It is
required that every one of the $2^{nR}$ possible messages should be sent
uncorrupted through the channel with high probability.}
\label{fig_Noisy_channel_coding}
\end{figure}

For the calculation, random coding will be used, that is $2^{nH\left(
X\right) }\ $strings will be chosen from the possible input strings, which
with high probability will belong in the set of typical ones. If these
strings are sent over the channel a message belonging in the set $Y^{n}$
will be received. But for each received letter $Y$ there is an ignorance on
knowing $X$ given by $H\left( Y|X\right) .$ Hence for each letter, $%
2^{H\left( Y|X\right) }$ bits could have been sent, which means that totally
there are $2^{nH\left( Y|X\right) }$ possible sent messages. In order to
achieve a reliable decoding the string received must be close to the $%
2^{nH\left( X\right) }$ initially chosen strings. Then decoding can be
modeled by drawing a Hamming sphere of radius $\delta $ around the received
message, containing $2^{n\left[ H\left( Y|X\right) +\delta \right] }$
possible input strings. In case exactly one input string belongs in this
sphere then the encoding-decoding scheme will work reliably. It is unlike
that no word will be contained in the Hamming sphere, but it should be
checked whether other input strings are contained in it. Each decoding
sphere contains a fraction%
\begin{equation*}
\frac{2^{n\left[ H\left( Y|X\right) +\delta \right] }}{2^{nH\left( X\right) }%
}=2^{-n\left[ H\left( Y:X\right) -\delta \right] },
\end{equation*}%
of the typical input strings. If there are $2^{nR}$ strings, where $R$ can
be related to Gilbert-Varshamov bound in equation \ref%
{Gilbert-Varshamov_bound}, the probability that one falls in the decoding
sphere by accident is%
\begin{equation*}
2^{nR}2^{-n\left[ H\left( Y:X\right) -\delta \right] }=2^{-n\left[ H\left(
Y:X\right) -R-\delta \right] }.
\end{equation*}%
Since $\delta $ can be chosen arbitrarily small, $R$ can be chosen to be as
close to $H\left( Y:X\right) $ as desired. Now getting the maximum over the
prior probabilities of the strings Shannon's result is found.

\begin{quote}
\textbf{Shannon's noisy channel coding theorem:}

For a noisy channel $\mathcal{N}$\ the capacity is given by%
\begin{equation*}
C\left( \mathcal{N}\right) =\underset{p\left( x\right) }{\max }\text{ }%
H\left( Y:X\right) ,
\end{equation*}%
where the maximum is taken over all input distributions $p\left( x\right) $ (%
\emph{a priori distributions}) for $X,$ for one use of the channel, and $\ Y$
is the corresponding induced random variable at the output of the channel.
\end{quote}

It should be noted that the capacity found in the above mentioned theorem is
the maximum one can get for the noisy channel $\mathcal{N}$.

\subsection{Classical information over noisy quantum channels\label%
{subsec_Class_over_noisy_Quant_ch}}

The case of sending classical information over noisy quantum channels is
quite similar to the classical channel. Each message is selected out of the $%
2^{nR},$ chosen by random coding as was done for the classical case. Suppose
now a message $M,$ is about to be sent and the $i$-th letter letter, denoted
by $M_{i}\in \left\{ 1,2,\ldots ,k\right\} ,$ is encoded in the quantum
states $\left\{ \rho _{1},\rho _{2},\ldots ,\rho _{k}\right\} $ of potential
inputs of a noisy channel represented by a quantum operation $\mathcal{E}$.
Then the message $M$ sent is written as a tensor product $\rho
_{M_{1}}\otimes \rho _{M_{2}}\otimes \cdots \otimes \rho _{M_{n}}.$ Because
of noise, the channel has some impact to the transmitted states, such that
the output states are $\sigma _{M_{i}}=\mathcal{E}\left( \rho
_{M_{i}}\right) ,$ thus the total impact on the message $M$ will be denoted $%
\sigma _{M}=\mathcal{E}^{\otimes n}\left( \rho _{M}\right) .$ The receiver
must decode the $\sigma _{M}$ message with a similar way to the one for the
noisy classical channel. Now because the channel is quantum, a set of POVM
measurements is going to describe the outcome of information on the part of
the receiver. To be more specific for every $M$ message a POVM operator $%
E_{M}$ is going to be corresponded. The probability of successfully decoding
the message, will be tr$\left( \sigma _{M}E_{M}\right) ,$ and therefore the
probability of an error being made for the message $M$ is $%
p_{M}^{e}\triangleq 1-$tr$\left( \sigma _{M}E_{M}\right) .$

The average probability of making an error while choosing from one of the $%
2^{nR}$ messages is%
\begin{equation}
p_{av}\triangleq \frac{\sum_{M}p_{M}^{e}}{2^{nR}}=\frac{\sum_{M}\left[ 1-%
\text{tr}\left( \sigma _{M}E_{M}\right) \right] }{2^{nR}}.
\label{average_probability_of_making_error}
\end{equation}%
Now the POVM operators $E_{M}$ can be constructed as follows. Let $\epsilon
>0,$ and assume $p_{j}$ is a probability distribution over the indices $%
\left\{ 1,2,\ldots ,k\right\} $ of the letters, named the \emph{a priori
distribution}, then for the space of output alphabet a matrix density can be
defined, $\bar{\sigma}\triangleq \sum_{j}p_{j}\sigma _{j},$ and let $P$ be a
projector onto the $\epsilon $-typical subspace of $\bar{\sigma}^{\otimes
n}. $ By the theorem of quantum typical sequences, it follows that for any $%
\delta >0$ and for sufficiently large $n,$ tr$\left( \bar{\sigma}^{\otimes
n}\left( I-P\right) \right) \leq \delta .$ For a given message $M$ the
notion of $\epsilon $-typical subspace for $\sigma _{M}$ can be defined,
based on the idea that typically $\sigma _{M}$ is a tensor of about $np_{1}$
copies of $\rho _{1},$ $np_{2}$ copies of $\rho _{2}$ and so on. Define $%
\bar{S}\triangleq \sum_{j}p_{j}S\left( \sigma _{j}\right) .$ Suppose $\sigma
_{j}$ has a spectral decomposition $\sum_{k}\lambda _{k}^{j}\left|
e_{k}^{j}\right\rangle \left\langle e_{k}^{j}\right| ,$ so $\sigma _{M}=%
\underset{K}{\sum }\lambda _{K}^{M}\left| E_{K}^{M}\right\rangle
\left\langle E_{K}^{M}\right| ,$ where $K=\left( K_{1},\ldots ,K_{n}\right)
, $ and for convenience $\lambda _{K}^{M}\triangleq \lambda
_{K_{1}}^{M_{1}}\lambda _{K_{2}}^{M_{2}}\cdots \lambda _{K_{n}}^{M_{n}}$ and 
$\left| E_{K}^{M}\right\rangle \triangleq \left|
e_{K_{1}}^{M_{1}}\right\rangle \left| e_{K_{2}}^{M_{2}}\right\rangle \cdots
\left| e_{K_{n}}^{M_{n}}\right\rangle $ are defined. Defining finally the
projector $P_{M}$ onto the space spanned by all $\left|
E_{K}^{M}\right\rangle $ such that%
\begin{equation}
\left| \frac{1}{n}\log \frac{1}{\lambda _{K}^{M}}-\bar{S}\right| \leq
\epsilon .  \label{def_of_the_projector_P_M}
\end{equation}%
Moreover the law of large numbers imply that for any $\delta >0$ and for
sufficiently large $n,$ $\mathbf{E}\left[ \text{tr}\left( \sigma
_{M}P_{M}\right) \right] \geq 1-\delta ,$ where the expectation is taken
with respect to the distribution over the strings $\rho _{M},$ hence $%
\mathbf{E}\left[ \text{tr}\left( \sigma _{M}\left( I-P_{M}\right) \right) %
\right] \leq \delta .$ Also note that by the definition (\ref%
{def_of_the_projector_P_M}) the dimension of the subspace onto which $P_{M}$
projects can be at most $2^{n\left( \bar{S}+\epsilon \right) },$ and thus $%
\mathbf{E}\left[ \text{tr}\left( P_{M}\right) \right] \leq 2^{n\left( \bar{S}%
+\epsilon \right) }.$ Now the POVM operators are defined%
\begin{equation}
E_{M}\triangleq \left( \underset{M^{\prime }}{\sum }PP_{M^{\prime }}P\right)
^{-\frac{1}{2}}PP_{M}P\left( \underset{M^{\prime }}{\sum }PP_{M^{\prime
}}P\right) ^{-\frac{1}{2}}.  \label{def_E_M}
\end{equation}%
To explain intuitively this construction, up to small corrections $E_{M}$ is
equal to the projector $P_{M}$ and the measurements $\left\{ E_{M}\right\} $
correspond essentially to checking whether the output of the channel falls
into the subspace on which $P_{M}$ projects. This can be though as analogous
to the Hamming sphere around the output. Using (\ref%
{average_probability_of_making_error}) and (\ref{def_data_proc_ineq}) one
can find out that $\mathbf{E}\left[ p_{av}\right] \leq 4\delta +\left(
2^{nR}-1\right) 2^{-n\left[ S\left( \bar{\sigma}\right) -\bar{S}-2\epsilon %
\right] }.$ Provided $R<S\left( \bar{\sigma}\right) -\bar{S}$ it follows
that $\mathbf{E}\left[ p_{av}\right] $ approaches zero as $n$ approaches
infinity. These where the main steps to prove the following theorem.

\begin{quotation}
\textbf{Holevo-Schumacher-Westmoreland (HSW) theorem:}

Let $\mathcal{E}$ be a trace-preserving quantum operation. Define%
\begin{equation}
\chi \left( \mathcal{E}\right) \triangleq \underset{\left\{ p_{j},\rho
_{j}\right\} }{\max }\left[ S\left( \mathcal{E}\left( \underset{j}{\sum }%
p_{j}\rho _{j}\right) \right) -\underset{j}{\sum }p_{j}S\left( \mathcal{E}%
\left( \rho _{j}\right) \right) \right] ,  \label{HSW_theorem}
\end{equation}%
where the maximum is over all ensembles $\left\{ p_{j},\rho _{j}\right\} $
of possible input states $\rho _{j}$ to the channel. Then $\chi \left( 
\mathcal{E}\right) $ is the product state capacity for then channel $%
\mathcal{E}$, that is, $\chi \left( \mathcal{E}\right) =C^{\left( 1\right)
}\left( \mathcal{E}\right) .$
\end{quotation}

In the aforementioned theorem the symbol $C^{\left( 1\right) }\left( 
\mathcal{E}\right) $ is used to denote the capacity of the channel, but just
in the case of a product case. Whether this kind of capacity might be
exceeded if the input states are prepared in entangled states is not known
and it is one of the many interesting open questions of quantum information
theory. It should be emphasized that like in the case of a classical
channel, the capacity found here is the maximum one can get for the noisy
channel $\mathcal{E}$.

Finally for the maximization in equation (\ref{HSW_theorem}) is potentially
over an unbounded set, therefore for practical reasons one takes the maximum
over and ensemble of pure states (refer to \cite[p.212-214]{Preskill} for
more details).

\subsection{Quantum information over noisy quantum channels\label%
{subsec_Quant_info_over_Qch}}

Unfortunately up to day there is no complete understanding of quantum
channel capacity. As far as it concerns the present state of knowledge, the
most important results were already presented in subsections \ref%
{subsec_Access_quant_info} and \ref{subsec_Quant_data_process}, concerning
each the accessible quantum information and quantum data processing. One
should also mention that there exist a quantum analogue to equation \ref%
{singleton_bound}, the \emph{quantum Singleton bound}, which is $n-k\geq
2\left( d-1\right) $ for an $\left[ n,k,d\right] $ quantum error correcting
code.

However an additional comment should be mentioned. The coherent information
defined in (\ref{def_coherent_info}), because of its r\^{o}le in quantum
data processing (equation (\ref{quantum_data_process}) compared with (\ref%
{def_data_proc_ineq})), it is believed to be the quantum analogue of mutual
information $H\left( X:Y\right) ,$ and hence perhaps related to quantum
channel capacity. This intuitive argument is yet unproven. For some progress
to that hypothesis, see \cite[p.606]{NielsenChuang} and the references
therein.

\chapter{Practical quantum cryptography\label{chap_Quantum_cryptography}}

One of the most interesting applications of quantum information theory and
the only one realized until now, is quantum cryptography. For an overview of
the history and the methods of classical cryptography, and for a simple
introduction in quantum cryptography refer to \cite{Singh}.

It is widely known that there exist many secure cryptographic systems, like
for example the RSA \cite{RSA}. Then why quantum cryptography is needed? The
reason is that as long quantum laws hold, it is theoretically unbreakable.
In addition to this all known classical cryptographic systems, like the RSA,
seem to be broken by quantum computers \cite{NielsenChuang,Singh}, using
quantum factoring and quantum discrete logarithms \cite{Shor94,Shor97}, or
by methods found in quantum search algorithms \cite{Grover96,Grover97}.

In this chapter the basic notions of quantum cryptography and a proof of its
security are analyzed in section \ref{sect_Quantum_Crypto}. Then the
possibility of constructing a commercial device capable of performing
quantum cryptography is discussed in section \ref%
{sect_Commercial_device_for_Quant_Crypto}.

\section{Theoretical principles of quantum cryptography\label%
{sect_Quantum_Crypto}}

\begin{figure}[tbh]
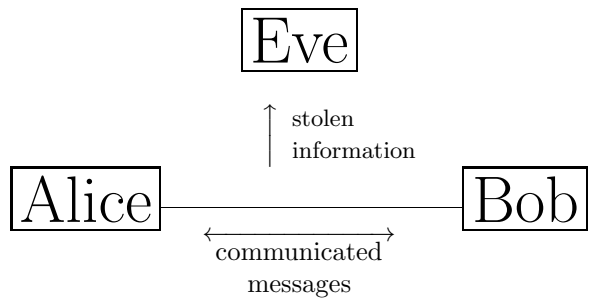

\begin{center}
\begin{tabular}{c}
\fbox{$\text{{\Huge Eve}}$} \\ 
\\ 
\hspace*{0.5in}$\left\uparrow 
\begin{array}{l}
\text{{\small stolen}} \\ 
\text{{\small information}}%
\end{array}%
\right. $ \\ 
\fbox{{\Huge Alice}}{\huge ------------------}\fbox{{\Huge Bob}} \\ 
$\overleftrightarrow{%
\begin{array}{c}
\text{communicated} \\ 
\text{messages}%
\end{array}%
}$%
\end{tabular}%
\end{center}
\caption{Alice and Bob communicating with the fear of Eve stealing
information.}
\label{fig_Alice-Bob_communication}
\end{figure}

Cryptography usually concerns two parties $A$ and $B,$ willing to securely
communicate, and possibly an eavesdropper $E.$ These points are sometimes
given human names for simplicity, calling $A:$ Alice, $B:$ Bob and $E:$ Eve.
The situation is visualized in Figure \ref{fig_Alice-Bob_communication}.

Quantum mechanics can be used to secure communication between Alice and Bob.
To achieve this a quantum line will be used to send a randomly produced
cryptographic key (see Figure \ref{fig_Experim_setup1}). This can be done
using protocols described in subsection \ref{subsec_BB84}. Moreover Alice
and Bob need a classical line to discuss their result, as described by the
same protocol, and send the encrypted message.

The encryption and the decryption of the message is done using a string $K,$
the key, which should be of equal length to the message, $M.$ Then by
applying the modulo $2$ addition $\oplus $ for each bit of the strings, the
encrypted message is $E=M\oplus K.$ Finally the message is decrypted by
further addition of the key, $\left( M\oplus K\right) \oplus K=M\oplus
\left( K\oplus K\right) =M\oplus 0=M.$

\begin{figure}[tbh]
\begin{center}
\includegraphics[width=2.8cm,height=8cm]{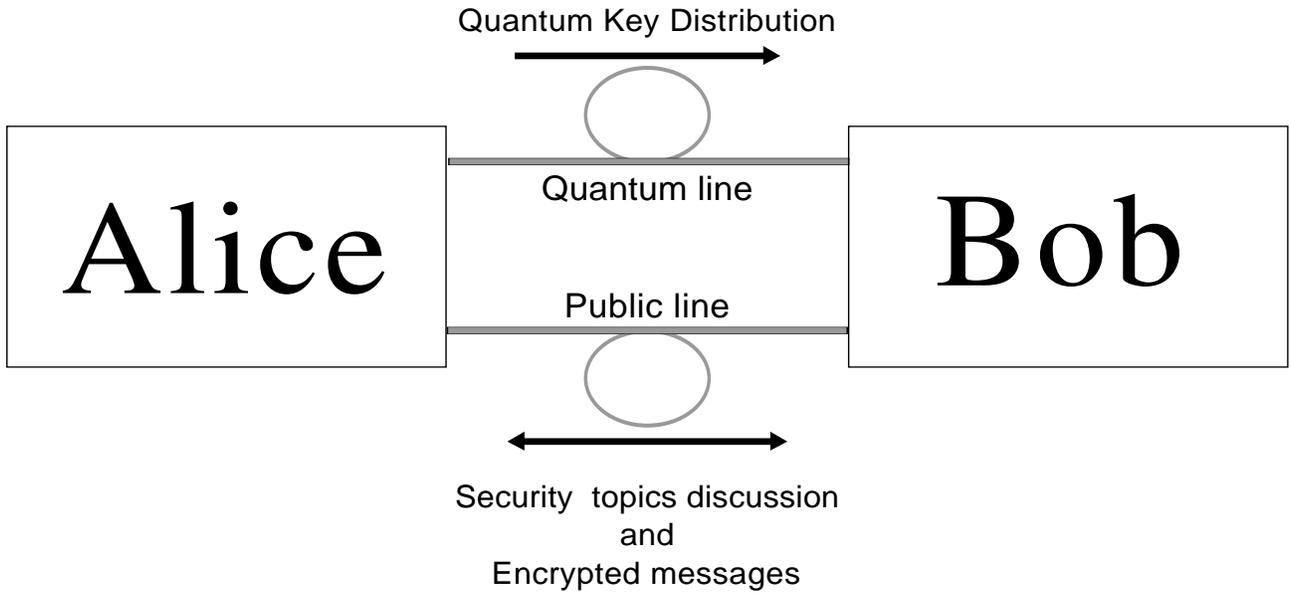}
\end{center}
\caption{The setup of BB84 quantum key distribution protocol. The quantum
line is used for the distribution of the key, while the classical line for
discussing security topics and the transmission of encrypted messages.}
\label{fig_QKD}
\end{figure}

\subsection{The BB84 quantum key distribution protocol\label{subsec_BB84}}

Looking back in Figure \ref{fig_Features_of_Quant_Mech}, presented in page %
\pageref{fig_Features_of_Quant_Mech}, one can see from the basic features of
quantum mechanics that it is not easy for Eve to steal information, since it
cannot be copied by no-cloning theorem. Moreover, it is impossible for her
to distinguish non-orthogonal states, and any information gain related to
such a distinction involves a disturbance. Then Alice and Bob by sacrificing
some piece of information, and checking through the coincidence of their key
can verify whether Eve was listening to them. Motivated by the above
arguments Bennet and Brassard presented in 1984 the first quantum
cryptographic protocol \cite{BennettBrassard} described next.

Suppose Alice has an $n$-bit message. Then in order to generate an equal
length cryptographic key she begins with some $a$ and $b$ $\left( 4+\delta
\right) n$-bit strings randomly produced. She must then encode this to a $%
\left( 4+\delta \right) n$-qubit string%
\begin{equation*}
\left| \psi \right\rangle \equiv \bigotimes_{k=1}^{\left( 4+\delta \right)
n}\left| \psi _{a_{k}b_{k}}\right\rangle ,
\end{equation*}%
where $a_{k}$ the $k$-th bit of $a$ (and similarly for $b$), and each qubit
just mentioned can be%
\begin{eqnarray}
\left| \psi _{00}\right\rangle &\triangleq &\left| 0\right\rangle ,
\label{BB84_def_4_states} \\
\left| \psi _{10}\right\rangle &\triangleq &\left| 1\right\rangle ,  \notag
\\
\left| \psi _{01}\right\rangle &\triangleq &\left| +\right\rangle =\frac{%
\left| 0\right\rangle +\left| 1\right\rangle }{\sqrt{2}},  \notag \\
\left| \psi _{11}\right\rangle &\triangleq &\left| -\right\rangle =\frac{%
\left| 0\right\rangle -\left| 1\right\rangle }{\sqrt{2}},  \notag
\end{eqnarray}%
where $\left| +\right\rangle $ is produced by application of the Hadamard
gate on $\left| 0\right\rangle ,$ and $\left| -\right\rangle $ by applying
the same gate on $\left| 1\right\rangle .$ The above procedure, encodes $a,$
in the basis $\left\{ \left| 0\right\rangle ,\left| 1\right\rangle \right\} $
if $b_{k}=0,$ or in $\left\{ \left| +\right\rangle ,\left| -\right\rangle
\right\} $ if $b_{k}=1.$ The states of each basis are not orthogonal to the
ones of the other basis, and hence they cannot be distinguished, without
distortion. $\delta $ is used as a tolerance due to noise of the channel.

The pure state $\left| \psi \right\rangle \left\langle \psi \right| $ is
sent over the quantum channel, and Bob receives $\mathcal{E}\left( \left|
\psi \right\rangle \left\langle \psi \right| \right) .$ He announces the
receipt in the public channel. At this stage Alice, Bob and Eve have each
their own states, described by their own density matrices. Moreover Alice
has not revealed $b$ during the transmission, and Eve has difficulty in
identifying $a$ by measurement, since she does not know in which basis she
should do it, and hence any trial to measure disturbs $\mathcal{E}\left(
\left| \psi \right\rangle \left\langle \psi \right| \right) .$ However $%
\mathcal{E}\left( \left| \psi \right\rangle \left\langle \psi \right|
\right) \neq \left| \psi \right\rangle \left\langle \psi \right| $ in
general, because the channel can be noisy. The above description implies
that $a$ and $b$ should be completely random, because any ability of Eve to
infer anything about these strings, would jeopardize the security of the
protocol. Note that quantum mechanics offers a way of perfect random number
generation, just by applying on $\left| 0\right\rangle $ state, the Hadamard
gate, one obtains $\frac{\left| 0\right\rangle +\left| 1\right\rangle }{%
\sqrt{2}},$ and after a measurement either $\left| 0\right\rangle $ or $%
\left| 1\right\rangle $ are returned with probability $p=\frac{1}{2}.$

In order to find their key Bob proceeds in measuring $\mathcal{E}\left(
\left| \psi \right\rangle \left\langle \psi \right| \right) $ in the basis
determined by a string $b^{\prime }$ he generated randomly too. This way he
measures the qubits and finds some string $a^{\prime }$. After the end of
his measurements he asks Alice, on the public channel, to inform him about $%
b.$ They then discard all bits $a_{m}$ and $a_{m}^{\prime }$ having $%
b_{m}\neq b_{m}^{\prime },$ since they measured them in different bases and
hence are uncorrelated. It should be stressed that the public announcement
of $b$ does not help Eve to infer anything about $a$ or $a^{\prime }.$

At this point they both have new keys $a_{\text{A}}^{\prime }$ and $a_{\text{%
B}}^{\prime }$ of statistically approximate length $2n,$ and Alice selects
randomly $n$-bits and informs Bob publicly about their values. If they find
that a number of qubits above a threshold $t,$ disagree then they stop the
procedure and retry. In case of many failures there is definitely an invader
and they should locate him. In case of success, there are some approximately 
$n$ bit strings $a_{\text{A}}^{\prime \prime }$ and $a_{\text{B}}^{\prime
\prime },$ not communicated in public. Then if for example Alice wants to
send a message $M,$ she encodes it, taking $E=M\oplus a_{\text{A}}^{\prime
\prime },$ and send $E$ through the public channel. Then Bob decodes it by $%
M^{\prime }=E\oplus a_{\text{B}}^{\prime \prime }.$ The current keys,
strings $a_{\text{A}}^{\prime \prime }$ and $a_{\text{B}}^{\prime \prime },$
should be discarded and not reused.

How can Alice's key $a_{\text{A}}^{\prime \prime }$ be the same as Bob's $a_{%
\text{B}}^{\prime \prime }$ in the case of a noisy channel which results $%
\mathcal{E}\left( \left| \psi \right\rangle \left\langle \psi \right|
\right) \neq \left| \psi \right\rangle \left\langle \psi \right| $? This is
the main topic of the next subsection.

\subsection{Information reconciliation and privacy amplification\label%
{subsec_Info_recon_and_priv_amplif}}

As anyone can assume, the communication over the noisy quantum channel is
imperfect, $\mathcal{E}\left( \left| \psi \right\rangle \left\langle \psi
\right| \right) \neq \left| \psi \right\rangle \left\langle \psi \right| $,
and hence even if there was no interference by Eve in general $a_{\text{A}%
}^{\prime }\neq a_{\text{B}}^{\prime }.$ Alice and Bob should perform an
error correction to get the same key $a^{\ast },$ by discussing over the
public channel and revealing some string $g$ related to $a_{\text{A}%
}^{\prime }$ and $a_{\text{B}}^{\prime }.$ This is named \emph{information
reconciliation}. In order to have a secure key they should have \emph{%
privacy amplification} by removing some bits of $a^{\ast }$ to get $%
a^{\prime \prime },$ minimizing Eve's knowledge, since $a_{\text{A}}^{\ast }$
are related to string $g$ publicly communicated. It is known that both
procedures can be used with high security.

As already seen information reconciliation is nothing more than
error-correction; it turns out that privacy amplification it related to
error-correction too, and both tasks are implemented using classical codes.
To be more specific decoding from a randomly chosen CSS code, already
presented in subsection \ref{subsec_Error_correction},\ can be thought of as
performing information reconciliation and privacy amplification. This can be
seen by considering their classical properties. Consider two classical
linear codes $C_{1}$ and $C_{2}$ which satisfy the conditions for a $t$ bit
error-correcting $\left[ n,m\right] $ CSS code. To perform the subsequent
task the channel should sometimes randomly tested and seen to have errors
less than $t,$ including eavesdropping.

Alice should pick randomly the codes $C_{1}$ and $C_{2}.$ Assume that $a_{%
\text{A}}^{\prime }=a_{\text{B}}^{\prime }+e,$ where $e$ is some error.
Since it is known that less than $t$ errors occurred, if Alice and Bob both
correct their states to the nearest codeword in $C_{1},$ their results will
be $a^{\ast }.$ This step is information reconciliation. To reduce Eve's
knowledge Alice and Bob identify which of the $2^{m}$ cosets of $C_{2}$ in $%
C_{1}$ their state $a^{\ast }$ belongs to. This is done by computing the
coset $a^{\ast }+C_{2}$ in $C_{1}.$ The result is their $m$ bit key sting $%
a^{\prime \prime }.$ By virtue of Eve's knowledge about $C_{2},$ and the
error-correcting properties of $C_{2}$, this procedure can reduce Eve's
mutual information with $a^{\prime \prime }$ to an acceptable level,
performing privacy amplification.

\subsection{Privacy and security}

In order to quantify bounds in quantum cryptography, two important notions
are defined in this subsection: privacy and security.

Assume Alice sends the quantum states $\rho _{k}^{A},$ $k=0,1,2,\ldots ,$
and Bob receives $\rho _{k}^{B}=\mathcal{E}\left( \rho _{k}^{A}\right) .$
The mutual information between the result of any measurement Bob may do and
Alice's value, $H\left( B:A\right) ,$ is bounded above by Holevo bound (\ref%
{Holevo_bound}), thus $H\left( B:A\right) \leq \chi ^{B},$ and similarly
Eve's mutual information is bounded above by $H\left( E:A\right) \leq \chi
^{E}.$ Since any excess information Bob has relative to Eve, at least above
a certain threshold can in principle be exploited by Bob and Alice to
distill a shared secret key using the techniques of the last subsection. It
makes sense to define \emph{privacy} as 
\begin{equation*}
\mathcal{P}\triangleq \underset{\mathcal{S}}{\sup }\left[ H\left( B:A\right)
-H\left( E:A\right) \right] ,
\end{equation*}%
where $\mathcal{S}$ are all the possible strategies Alice and Bob may employ
to the channel. This is the maximum excess classical information that Bob
may obtain relative to Eve about Alice's quantum signal. Using the HSW
theorem (\ref{HSW_theorem}), Alice and Bob may employ a strategy such that $%
H\left( B:A\right) =\chi ^{B},$ while for any strategy Eve may employ, $%
H\left( E:A\right) \leq \chi ^{E}.$ Thus $\mathcal{P}\geq \chi ^{B}-\chi
^{E}.$ A lower bound may be obtained by assuming that Alice's signal states
are pure (refer to discussion after equation (\ref{HSW_theorem}) to see
why), $\rho ^{A}=\left| \psi _{k}^{A}\right\rangle \left\langle \psi
_{k}^{A}\right| ,$ and if Eve initially had an unentangled state $\left|
0^{E}\right\rangle ,$ which may also assumed pure. Suppose Eve's interaction
is $\left| \psi ^{EB}\right\rangle =U\left| \psi _{k}^{A}\right\rangle
\left| 0^{E}\right\rangle ,$ since it is a pure state, the reduced density
matrices $\rho _{k}^{E}$ and $\rho _{k}^{B}$ will have the same non-zero
eigenvalues, and thus the same entropies, $S\left( \rho _{k}^{E}\right)
=S\left( \rho _{k}^{B}\right) .$ Thus $\mathcal{P}\geq \chi ^{B}-\chi
=S\left( \rho ^{B}\right) -\sum_{k}p_{k}S\left( \rho _{k}^{B}\right)
-S\left( \rho ^{E}\right) +\sum_{k}p_{k}S\left( \rho _{k}^{E}\right)
=S\left( \rho ^{B}\right) -S\left( \rho ^{E}\right) =I\left( \rho ,\mathcal{E%
}\right) ,$ where the definition (\ref{def_coherent_info}) was used. Note
that the lower bound for privacy is protocol independent.

A quantum key distribution (QKD) protocol is defined \emph{secure} if, for
any security parameters $s>0$ and $l>0$ chosen by Alice and Bob, and for any
eavesdropping strategy, either the scheme aborts, or succeeds with
probability at least $1-O\left( 2^{-s}\right) ,$ and guarantees that Eve's
mutual information with the final key is less than $2^{-l}.$ The key string
must also be essentially random.

\subsection{A provable secure version of the BB84 protocol\label%
{subsec_Secure_BB84}}

It can be proven \cite[p.593-599]{NielsenChuang} that using CSS codes one
can have a 100\% secure quantum key distribution. However CSS codes need
perfect quantum computation, which is not yet achieved. Fortunately there is
a chance of using the classical properties of CSS$_{u,v}$ codes defined in (%
\ref{CSS_quant_error_corr_used_in_QKD}) to have an equally secure classical
computation version of BB84 protocol (refer to \cite[p.599-602]%
{NielsenChuang} for a proof). This version is made up of the following steps:

\begin{enumerate}
\item Alice creates $\left( 4+\delta \right) n$ random bits.

\item For each bit, she creates a qubit either in the $\left\{ \left|
0\right\rangle ,\left| 1\right\rangle \right\} $ or in $\left\{ \left|
+\right\rangle ,\left| -\right\rangle \right\} $ basis, according to a
random bit string $b,$ see for example (\ref{BB84_def_4_states}).

\item Alice sends the resulting qubits to Bob.

\item Alice chooses a random $v_{k}\in C_{1}.$

\item Bob receives the qubits, publicly announces this fact, and measures
each in the $\left\{ \left| 0\right\rangle ,\left| 1\right\rangle \right\} $
or in $\left\{ \left| +\right\rangle ,\left| -\right\rangle \right\} $ basis
randomly.

\item Alice announces $b.$

\item Alice and Bob discard those bits Bob measured in a basis other than $%
b. $ With high probability, there are at least $2n$ bits left; if not, abort
the protocol. Alice decides randomly on a set of $2n$ bits to continue to
use, randomly selects $n$ of these to check bits, and announces the
selection.

\item Alice and Bob publicly compare their check bits. If more than $t$ of
these disagree, they abort the protocol. Alice is left with the $n$ bit
string $x,$ and Bob with $x+\epsilon .$

\item Alice announces $x-v_{k}.$ Bob subtracts this from his result,
correcting it with code $C_{1}$ to obtain $v_{k}.$

\item Alice and Bob compute the coset of $v_{k}+C_{2}$ in $C_{1}$ to obtain
the key $k.$
\end{enumerate}

\section{A commercial quantum cryptographic device\label%
{sect_Commercial_device_for_Quant_Crypto}}

After a theoretical presentation of quantum key distribution it is time to
present experimental results and then discuss the possibility of having a
commercial device realizing quantum cryptography. These two topics are
discussed correspondingly in subsections \ref{subsec_Exper_QKD} and \ref%
{subsec_Commercial_Quant_Crypto}.

\subsection{Experimental quantum key distribution\label{subsec_Exper_QKD}}

The first demonstration of quantum key distribution was performed at the IBM
laboratory in 1989 \cite{Bennet-et-al}\ over a distance of 30 cm. Since then
there has been remarkable improvement, demonstrating quantum key
distribution over a distance of 10 km \cite{BR98,BR00}, in IBM too, or over
distances exceeding 40km, and also in installed telecommunication fiber,
under the Lake Geneva \cite{MZG96}.

In most cases experimental quantum cryptography is done using single
photons, and optical fibers are used to guide them from Alice to Bob. Once
the medium is chosen one should pick the right source and detectors. Since
they have to be compatible, the crucial choice is the wave length. There are
two main possibilities. Either one chooses a wavelength around 800 nm where
efficient photon counters are commercially available, or one chooses a
wavelength compatible with today's telecommunication optical fibers, that is
near 1300 nm or 1550 nm. The first choice requires the use of special
fibers, hence the installed telecommunications networks can't be used. The
second choice requires the improvement or development of new detectors not
based on silicon semiconductors which are transparent above 1000 nm
wavelength. It is still unclear which of the two alternatives will turn out
to be the best choice.

In what concerns the production of photons according to the BB84 states,
defined in equation (\ref{BB84_def_4_states}), one can choose different
polarization states, which form non-orthogonal bases. Polarization can be
for example linear, identifying $\left| 0\right\rangle \equiv \left|
\uparrow \right\rangle $ and $\left| 1\right\rangle \equiv \left| \downarrow
\right\rangle ,$ or circular identifying $\left| +\right\rangle \equiv
\left| \circlearrowleft \right\rangle $ and $\left| -\right\rangle \equiv
\left| \circlearrowright \right\rangle .$ However in practice single photon
states are difficult to realize thus approximately single photon states are
produced by faint laser pulses.

Finally one should detect these approximate single photon states. This is
achieved using a variety of techniques. One can choose between
photomultipliers, avalanche-photodiodes, multichannel plates and
supraconducting Josephson junctions.

\begin{figure}[tbh]
\begin{center}
\includegraphics[width=6.4cm,height=7cm]{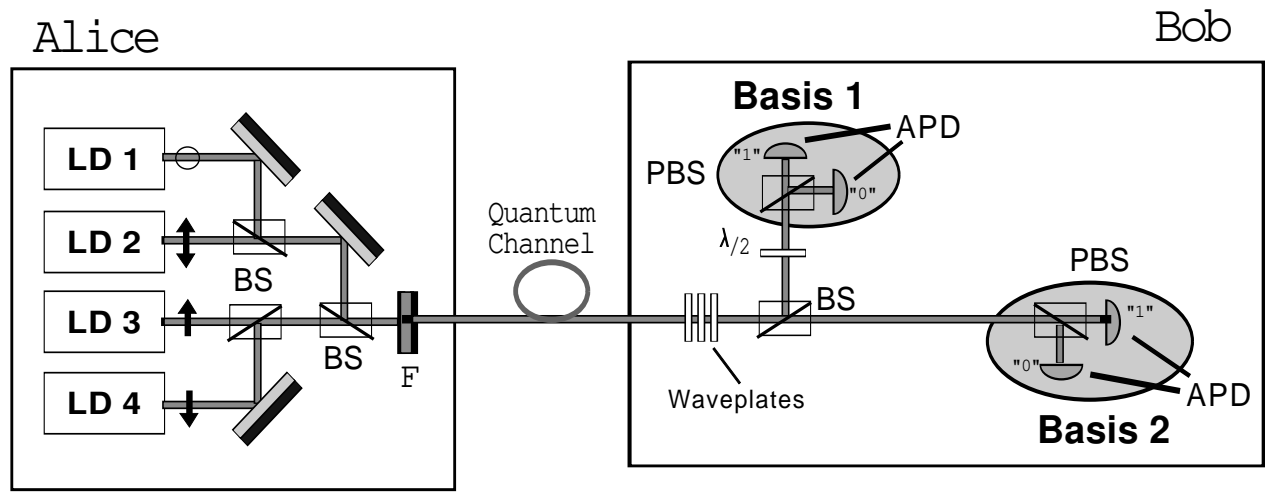}
\end{center}
\caption{Typical system for quantum cryptography using polarization coding
(LD: laser diode, BS: beamsplitter, F: neutral density filter, PBS:
polarizing beam splitter, $\protect\lambda /2$: half waveplate, APD:
avalanche photodiode).}
\label{fig_Experim_setup1}
\end{figure}

A typical experimental setup for quantum key distribution, with the
technology described above is sawn in Figure \ref{fig_Experim_setup1}.

For more experimental details on the single photon quantum key distribution
one should refer to \cite[p.11-29]{Gisin-et-al}.

\subsection{Commercial quantum cryptography\label%
{subsec_Commercial_Quant_Crypto}}

Once the experimental setup of quantum key distribution has been performed,
as discussed in subsection \ref{subsec_Exper_QKD} and as sawn in Figures \ref%
{fig_Experim_setup1} and \ref{fig_QKD}, one needs to follow the steps
presented in subsection \ref{subsec_Secure_BB84}, in order to achieve
perfectly secure quantum cryptography. Of course these steps are nothing but
an algorithm, which can be analyzed and easily implemented into a program
which can run on current technology computers. Moreover special controlling
devices are needed in order to instruct the laser diodes, the beam splitters
and get the result of the measurements from the avalanche photodiodes. Such
hardware is already available in the market. One should not forget that a
public connection is needed, like for example the internet. For this reason
there exist patents \cite{Bakopoulos-et-al}, which can convert quantum
cryptography from a laboratory experiment into a commercial product. Such a
patent is discussed bellow.

One can reconstruct the steps of subsection \ref{subsec_Secure_BB84} into
the following computer program:

\begin{enumerate}
\item Alice's computer creates $\left( 4+\delta \right) n$ random bits,
using an unknown to anybody else algorithm. (One can also use quantum
techniques as discussed in the third paragraph of subsection \ref%
{subsec_BB84})

\item For each bit, Alice's computer triggers with a controller one of the
four laser diodes, as shown in Figure \ref{fig_QKD}, according to a random
bit string $b.$

\item This way the light beams are sent to Bob's site into an agreed
beforehand rate.

\item Alice's computer has already implemented the classical version of CSS
codes, and then it picks up randomly a $v_{k}\in C_{1}.$

\item Bob's computer instructs with a controller the beam splitter sawn in
Figure \ref{fig_QKD} in which basis to measure the light beam. The selection
should be according to a random algorithm. There should be a device
returning the result of the measurement of the avalanche photodiodes to
Bob's computer. Bob's computer announces the fact through an internet line
to Alice's computer.

\item Alice's computer announces $b.$

\item Both computers discard those bits Bob's measured in a basis other than 
$b.$ With high probability, there are at least $2n$ bits left; if not, abort
the protocol. Alice's computer decides randomly on a set of $2n$ bits to
continue to use, randomly selects $n$ of these to check bits, and announces
to Bob's computer through internet.

\item Both computers compare through internet their check bits. If more than 
$t$ of these disagree, they abort the protocol. Alice's computer is left
with the $n$ bit string $x,$ and Bob's with $x+\epsilon .$

\item Alice's computer announces $x-v_{k}.$ Bob's computer subtracts this
from his result, correcting it with code $C_{1}$ to obtain $v_{k}.$

\item Both computer calculate the coset of $v_{k}+C_{2}$ in $C_{1}$ to
obtain the key $k.$
\end{enumerate}

Concluding it should be noted that what is very important about the above
implementation, is that it is completely automatic and almost no human
intervention is needed. Thus users can just write their messages, command
the computers to send them securely and in the other side receive them.
Automatic process is what makes such a device successfully commercial.

\appendix

\chapter*{Summary\label{chap_Summary}}

\setcounter{chapter}{19} 
\addtocontents{toc}
{\contentsline {chapter}{Summary}
{\pageref{chap_Summary}}} \setcounter{figure}{0}

The most important tools and results of classical and quantum information
theory, obtained in the present Individual Study Option, are summarized in
the Figure \ref{fig_Sum_Info_Theory}.

\begin{figure}[tbh]
\begin{center}
\fbox{%
\begin{tabular}{c}
{\huge Information Theory} \\ \hline
\\ 
\begin{tabular}{c|c}
{\Large Classical} & \hspace*{0.23in}{\Large Quantum}\hspace*{0.23in} \\ 
&  \\ 
$\text{Shannon entropy}$ & \hspace*{0.23in}Von Neumann
entropy\hspace*{0.23in} \\ 
$H\left( X\right) =-\sum_{x}p_{x}\log p_{x}$ & \hspace*{0.23in}$S\left( \rho
\right) =-$tr$\left( \rho \log \rho \right) $\hspace*{0.23in}%
\end{tabular}
\\ 
\\ 
\textbf{Distinguishability and accessible information} \\ 
\begin{tabular}{c|c}
Letters always distinguishable & \hspace*{0.14in}Holevo bound\hspace*{0.14in}
\\ 
$N=\left| X\right| $ & \hspace*{0.14in}$H\left( X:Y\right) \leq S\left( \rho
\right) -\underset{x}{\sum }p_{x}S\left( \rho _{x}\right) $\hspace*{0.14in}
\\ 
& \hspace*{0.14in}$\rho =\underset{x}{\sum }p_{x}\left( \rho _{x}\right) $%
\hspace*{0.14in}%
\end{tabular}
\\ 
\\ 
\textbf{Information-theoretic relations} \\ 
\begin{tabular}{c|c}
Fano inequality & \hspace*{0.06in}Quantum Fano inequality\hspace*{0.06in} \\ 
$H\left( p_{e}\right) +p_{e}\log \left( \left| X\right| -1\right) \geq
H\left( X|Y\right) $ & \hspace*{0.06in}$H\left( F\left( \rho ,E\right)
\right) +\left( 1-F\left( \rho ,E\right) \right) \log \left( d^{2}-1\right) $%
\hspace*{0.06in} \\ 
& \hspace*{0.06in}$\geq S\left( \rho ,E\right) $\hspace*{0.06in} \\ 
Mutual information & \hspace*{0.06in}Coherent information\hspace*{0.06in} \\ 
$H\left( X:Y\right) =H\left( Y\right) -H\left( Y|X\right) $ & 
\hspace*{0.06in}$I\left( \rho ,E\right) =S\left( E\left( \rho \right)
\right) -S\left( \rho ,E\right) $\hspace*{0.06in} \\ 
Data processing inequality & \hspace*{0.06in}Quantum data processing
inequality\hspace*{0.06in} \\ 
$X\rightarrow Y\rightarrow Z$ & \hspace*{0.06in}$\rho \rightarrow \mathcal{E}%
_{1}\left( \rho \right) \rightarrow \left( \mathcal{E}_{2}\circ \mathcal{E}%
_{1}\right) \left( \rho \right) $\hspace*{0.06in} \\ 
$H\left( X\right) \geq H\left( X:Y\right) \geq H\left( X:Z\right) $ & 
\hspace*{0.06in}$S\left( \rho \right) \geq I\left( \rho ,\mathcal{E}%
_{1}\right) \geq I\left( \rho ,\mathcal{E}_{2}\circ \mathcal{E}_{1}\right) $%
\hspace*{0.06in}%
\end{tabular}
\\ 
\\ 
\textbf{Noiseless channel coding} \\ 
\begin{tabular}{c|c}
Shannon's theorem & \hspace*{0.09in}Schumacher's theorem\hspace*{0.09in} \\ 
$n_{\text{bits}}=H\left( X\right) $ & \hspace*{0.09in}$n_{\text{qubits}%
}=S\left( \sum_{x}p_{x}\rho _{x}\right) $\hspace*{0.09in}%
\end{tabular}
\\ 
\\ 
\textbf{Capacity of noisy channels for classical information} \\ 
\begin{tabular}{c|c}
Shannon's noisy coding theorem & Holevo-Schumacher-Westmoreland \\ 
& theorem \\ 
$C\left( \mathcal{N}\right) =\underset{p\left( x\right) }{\max }\text{ }%
H\left( Y:X\right) $ & $C^{\left( 1\right) }\left( \mathcal{E}\right) =%
\underset{\left\{ p_{j},\rho _{j}\right\} }{\max }\left[ S\left( \rho
^{\prime }\right) -\underset{x}{\sum }p_{x}S\left( \rho _{x}^{\prime
}\right) \right] $ \\ 
& $\rho _{x}^{\prime }=\mathcal{E}\left( \rho _{x}\right) ,$ $\rho ^{\prime
}=\underset{x}{\sum }p_{x}\rho _{x}^{\prime }$%
\end{tabular}%
\end{tabular}%
}
\end{center}
\caption{Summary of classical and quantum information theory.}
\label{fig_Sum_Info_Theory}
\end{figure}

The most important results concerning quantum cryptography, are summarized
in Figure (\ref{fig_Sum_Quant_Crypto}).

\begin{figure}[tbh]
\begin{center}
\fbox{%
\begin{tabular}{c}
{\huge Quantum Cryptography} \\ \hline
\\ 
\textbf{BB84 protocol, sends the following states} \\ 
$\bigotimes_{k=1}^{\left( 4+\delta \right) n}\left| \psi
_{a_{k}b_{k}}\right\rangle $ \\ 
see (\ref{BB84_def_4_states}) for definitions \\ 
\\ 
\textbf{Privacy of a quantum channel }$\mathcal{E}$ \\ 
$\mathcal{P}\geq I\left( \rho ,\mathcal{E}\right) $%
\end{tabular}%
}
\end{center}
\caption{Summary of quantum cryptography.}
\label{fig_Sum_Quant_Crypto}
\end{figure}

\addtocontents{toc}
{\contentsline {chapter}{Appendices}
{\pageref{appendix_Special_diagonal_normal_matr}}} \setcounter{chapter}{0}

\chapter{Omitted proofs}

\section{Special diagonal transformation of normal matrices\label%
{appendix_Special_diagonal_normal_matr}}

In this appendix it is going to be demonstrated that for a $d\times d$
normal matrix $A$, there exists a set of unitary matrices $U_{i}^{\left(
A\right) },$ such that $\sum_{i=1}^{d}U_{i}^{\left( A\right) }AU_{i}^{\left(
A\right) \dagger }=$tr$\left( A\right) I.$ Since $A$ is a normal matrix, by
spectral decomposition theorem there exists a matrix $U^{\left( A\right) }$
such that $U^{\left( A\right) }AU^{\left( A\right) \dagger }$ is diagonal,
and for this diagonal matrix there exists a unitary transformation $%
V_{i\leftrightarrow j}AV_{i\leftrightarrow j}^{\dagger }$ which interchanges
the $i$-th diagonal element with the $j$-th.

To see this suppose $B$ is a $d\times d$ dimensional diagonal matrix, then
the following matrix%
\begin{equation*}
\left( V_{i\leftrightarrow j}\right) _{kl}\triangleq \left\{ 
\begin{array}{c}
\delta _{kl},\text{ }k\text{ or }l\neq i,j \\ 
1,\text{ }k=j\text{ and }l=i \\ 
1,\text{ }k=i\text{ and }l=j \\ 
0,\text{ else}%
\end{array}%
\right. ,
\end{equation*}%
can be used to exchange the $i$-th diagonal element with the $j$-th of $B.$
Following the next steps $V_{i\leftrightarrow j}$ is initially proven to be
unitary%
\begin{eqnarray*}
V_{i\leftrightarrow j}V_{i\leftrightarrow j}^{\dagger }
&=&\sum_{l=1}^{d}\left( V_{i\leftrightarrow j}\right) _{kl}\left(
V_{i\leftrightarrow j}^{\dagger }\right) _{ln}=\sum_{l=1}^{d}\left(
V_{i\leftrightarrow j}\right) _{kl}\left( V_{i\leftrightarrow j}\right) _{nl}
\\
&=&\sum_{l=1}^{d}\left( 
\begin{array}{c}
\delta _{kl},\text{ }k\text{ or }l\neq i,j \\ 
1,\text{ }k=j\text{ and }l=i \\ 
1,\text{ }k=i\text{ and }l=j \\ 
0,\text{ else}%
\end{array}%
\right) \left( 
\begin{array}{c}
\delta _{nl},\text{ }n\text{ or }l\neq i,j \\ 
1,\text{ }n=j\text{ and }l=i \\ 
1,\text{ }n=i\text{ and }l=j \\ 
0,\text{ else}%
\end{array}%
\right) \\
&=&\left( 
\begin{array}{c}
\delta _{kn},\text{ }k\text{ or }n\neq i,j \\ 
1,\text{ }k=j\text{ and }n=j \\ 
1,\text{ }k=i\text{ and }n=i \\ 
0,\text{ else}%
\end{array}%
\right) =\delta _{kn}=I.
\end{eqnarray*}%
The ability of $V_{i\leftrightarrow j}$ to exchange the diagonal elements of
matrix $B=diag\left\{ \ldots ,b_{i},\ldots ,b_{j},\ldots \right\} $ is
exhibited by%
\begin{eqnarray*}
V_{i\leftrightarrow j}BV_{i\leftrightarrow j}^{\dagger }
&=&\sum_{l=1}^{d}\sum_{m=1}^{d}\left( V_{i\leftrightarrow j}\right)
_{kl}B_{lm}\left( V_{i\leftrightarrow j}^{\dagger }\right)
_{mn}=\sum_{l=1}^{d}\left( V_{i\leftrightarrow j}\right) _{kl}b_{l}\left(
V_{i\leftrightarrow j}^{\dagger }\right) _{nl} \\
&=&\sum_{l=1}^{d}\left( 
\begin{array}{c}
\delta _{kl},\text{ }k\text{ or }l\neq i,j \\ 
1,\text{ }k=j\text{ and }l=i \\ 
1,\text{ }k=i\text{ and }l=j \\ 
0,\text{ else}%
\end{array}%
\right) b_{l}\left( 
\begin{array}{c}
\delta _{nl},\text{ }n\text{ or }l\neq i,j \\ 
1,\text{ }n=j\text{ and }l=i \\ 
1,\text{ }n=i\text{ and }l=j \\ 
0,\text{ else}%
\end{array}%
\right) \\
&=&\left( 
\begin{array}{c}
b_{n}\delta _{kn},\text{ }k\neq i,j\text{ or }n\neq i,j \\ 
b_{i},\text{ }k=j\text{ and }n=j \\ 
b_{j},\text{ }k=i\text{ and }n=i \\ 
0,\text{ else}%
\end{array}%
\right) =diag\left\{ \ldots ,b_{j},\ldots ,b_{i},\ldots \right\} .
\end{eqnarray*}

Defining now $V_{23\cdots d1}\triangleq V_{1\leftrightarrow
2}V_{2\leftrightarrow 3}\cdots V_{d-1\leftrightarrow d}V_{d\leftrightarrow
1},$ which is unitary matrix as a multiplication of unitary matrices, then $%
V_{23\cdots d1}U^{\left( A\right) }AU^{\left( A\right) \dagger }V_{23\cdots
d1}^{\dagger }$ is a diagonal matrix where the in the first diagonal place
is the second diagonal element of $A$, in the second diagonal place the
third diagonal element of $A$, and so on until the final diagonal place
where the first diagonal element of $A$ stands. The next display visualizes
the similarity transformation $V_{23\cdots d1}$%
\begin{equation*}
\left[ 
\begin{array}{ccccccc}
a_{11} &  &  &  &  &  &  \\ 
& a_{22} &  &  &  &  &  \\ 
&  &  & \ddots &  &  &  \\ 
&  &  &  & a_{\left( d-2\right) \left( d-2\right) } &  &  \\ 
&  &  &  &  & a_{\left( d-1\right) \left( d-1\right) } &  \\ 
&  &  &  &  &  & a_{dd}%
\end{array}%
\right] \overset{V_{23\cdots d1}}{\rightarrow }\left[ 
\begin{array}{ccccccc}
a_{22} &  &  &  &  &  &  \\ 
& a_{33} &  &  &  &  &  \\ 
&  &  & \ddots &  &  &  \\ 
&  &  &  & a_{\left( d-1\right) \left( d-1\right) } &  &  \\ 
&  &  &  &  & a_{dd} &  \\ 
&  &  &  &  &  & a_{11}%
\end{array}%
\right] .
\end{equation*}%
Enumerating all the cyclic permutations of $\left( 1,2,\ldots ,d\right) $
with a number $i$ then the following unitary matrices are defined $%
U_{i}^{\left( A\right) }\triangleq V_{i}U^{\left( A\right) }.$ It is
straightforward to verify that,%
\begin{eqnarray*}
\sum_{i=1}^{d}U_{i}^{\left( A\right) }AU_{i}^{\left( A\right) \dagger } &=& 
\left[ 
\begin{array}{cccc}
a_{11}+a_{22}+\ldots +a_{dd} &  &  &  \\ 
& a_{22}+a_{33}+\ldots +a_{11} &  &  \\ 
&  & \ddots &  \\ 
&  &  & a_{dd}+a_{11}+\ldots +a_{\left( d-1\right) \left( d-1\right) }%
\end{array}%
\right] \\
&=&\text{tr}\left( A\right) I.
\end{eqnarray*}

\section{Projective measurements are equivalent to an addition of unitary
operations\label{appendix_Proj_measur_and_Unitary}}

Let $P$ be a projector and $Q\triangleq I-P$ the complementary projector, in
this appendix it will be sawn that there exist unitary matrices $U_{1},$ $%
U_{2}$ and a probability $p$ such that for all $\rho $, $P\rho P+Q\rho
Q=pU_{1}\rho U_{1}^{\dagger }+\left( 1-p\right) U_{2}\rho U_{2}^{\dagger }.$
Choose $p\triangleq \frac{1}{2},$ $U_{1}\triangleq Q-P$ and $U_{2}\triangleq
Q+P\triangleq I.$ It is obvious that $U_{1}U_{1}^{\dagger }=\left(
Q-P\right) \left( Q-P\right) =QQ+QP-PQ-+PP=Q+0+P=I,$ and $%
U_{2}U_{2}^{\dagger }=II=I.$ Now it is easy to check that 
\begin{eqnarray*}
\frac{1}{2}U_{1}\rho U_{1}^{\dagger }+\frac{1}{2}U_{2}\rho U_{2}^{\dagger }
&=&\frac{1}{2}\left( Q-P\right) \rho \left( Q-P\right) +\frac{1}{2}\left(
Q+P\right) \rho \left( Q+P\right) \\
&=&\frac{1}{2}\left( Q\rho Q-Q\rho P-P\rho Q+P\rho P\right) +\frac{1}{2}%
\left( Q\rho Q+Q\rho P+P\rho Q+P\rho P\right) \\
&=&P\rho P+Q\rho Q.
\end{eqnarray*}%
q.e.d. (Abbas Edalat contributed to this proof)

\chapter{Distance measures of information}

\section{Classical information distance: Hamming\label{appendix_Hamming}}

The \emph{Hamming distance} of two strings, is defined to be the number of
places their bits are different. Assuming $a$ and $b$ are the two $n$-bit
strings, and $a_{i}$ and $b_{i}$ are their $i$-th bits, one can write%
\begin{equation*}
d\left( a,b\right) \triangleq \sum_{i=1}^{n}\left| a_{i}-b_{i}\right| .
\end{equation*}%
Very naturally, a \emph{Hamming sphere} of center $c$ and radius $\delta ,$
is defined as the set of stings which have distance from $c$ less or equal
to $\delta $%
\begin{equation*}
\text{Sph}\left( c,\delta \right) \triangleq \left\{ s:d\left( c,s\right)
\leq \delta \right\} .
\end{equation*}

\section{Quantum information distance: Fidelity\label{appendix_Fidelity}}

Fidelity is a measure of distance of two quantum states, defined by%
\begin{equation*}
F\left( \rho ,\sigma \right) \triangleq \text{tr}\sqrt{\rho ^{\frac{1}{2}%
}\sigma \rho ^{\frac{1}{2}}}=\underset{\left| \psi \right\rangle ,\left|
\phi \right\rangle }{\max }\text{ }\left| \left\langle \psi |\phi
\right\rangle \right| ,
\end{equation*}%
and used to determine how well a quantum channel preserves information. To
do this the following function can be defined%
\begin{equation*}
F_{\text{min}}\left( \mathcal{E}\right) \triangleq \underset{\left| \psi
\right\rangle }{\min }\text{ }F\left( \left| \psi \right\rangle ,\mathcal{E}%
\left( \left| \psi \right\rangle \left\langle \psi \right| \right) \right) .
\end{equation*}%
where the quantum channel is simulated via the quantum operation $\mathcal{E}
$, and the minimization is considered as the worst case of a quantum signal.
Another interesting definition is%
\begin{equation*}
\bar{F}\triangleq \sum_{j}p_{j}F\left( \rho _{j},\mathcal{E}\left( \rho
_{j}\right) \right) ^{2},
\end{equation*}%
named the \emph{ensemble average fidelity}. Finally it is important to
quantify, how much entanglement between $R$ and $Q,$ sent through a quantum
channel $\mathcal{E}$ (a trace preserving operation), is preserved. This can
be done by \emph{entanglement fidelity}%
\begin{equation}
F\left( \rho ,\mathcal{E}\right) \triangleq F\left( RQ,R^{\prime }Q^{\prime
}\right) ^{2}=\left\langle RQ\right| \left[ \left( \mathcal{I}_{R}\otimes 
\mathcal{E}\right) \left( \left| RQ\right\rangle \left\langle RQ\right|
\right) \right] \left| RQ\right\rangle =\sum_{j}\left| \text{tr}\left( \rho
E_{i}\right) \right| ^{2},  \label{fidelity_entangl}
\end{equation}%
where the primes are for the states after the application of $\mathcal{E}$,
and $E_{i}$ are the operation elements of $\mathcal{E}$.

\end{document}